\documentclass[12pt]{iopart}
\usepackage{amscd}
\usepackage{bbm}
\usepackage{xy}
\usepackage{amssymb}
\usepackage{graphicx}
\usepackage{theorem}
\usepackage{makeidx}
\usepackage{epic}
\usepackage{epsfig}
\usepackage{times}
\usepackage{verbatim}
\usepackage[T1]{fontenc}
\usepackage{color}

\newtheorem{Theorem}{Theorem}[section]
\newtheorem{Prop}[Theorem]{Prop.}

\begin{document}

\title{Analytic structure of solutions of the one-dimensional Burgers equation with modified dissipation}
\author{Walter Pauls$^{1,2}$ and Samriddhi Sankar Ray$^3$ \footnote{Author to whom all correspondence should be addressed.}}

\address{
$^1$Max-Planck-Institut f\"ur Dynamik und Selbstorganisation, 
Am Fa{\ss}berg 11, 37073 G\"ottingen, Germany\\
$^2$ M\"uhlweg 4, 73460 H\"uttlingen, Germany\\ 
$^3$International Centre for
  Theoretical Sciences, Tata Institute of Fundamental Research,
  Bangalore 560089, India}
\date{}

\begin{abstract}

We use the one-dimensional Burgers equation to illustrate the effect of
replacing the standard Laplacian dissipation term by a more general
function of the Laplacian -- of which hyperviscosity 
is the best known example -- in equations of hydrodynamics.  We
analyze the asymptotic structure of solutions in the Fourier space at very high
wave-numbers by introducing an approach applicable to a wide class of
hydrodynamical equations whose solutions are calculated in the limit of
vanishing Reynolds numbers from algebraic recursion relations involving
iterated integrations. We give a detailed analysis of their analytic structure
for two different types of dissipation: a  hyperviscous and an exponentially
growing dissipation term.  Our results, obtained in the limit of vanishing
Reynolds numbers, are validated by high-precision numerical simulations at
non-zero Reynolds numbers.  We then study the bottleneck problem, an
intermediate asymptotics phenomenon, which in the case of the Burgers equation
arises when ones uses dissipation terms  (such as hyperviscosity) growing
faster at high wave-numbers than the standard Laplacian dissipation term. A
linearized solution of  the well-known boundary layer limit of the Burgers
equation involving two numerically determined parameters gives a good
description of the bottleneck region.  

\end{abstract}

\section{Introduction}
The physics of a fluid in a turbulent state is multiscale. Hence, it is 
convenient to study turbulence by separating the scales into energy injection
$L$, inertial $r$, and dissipation $\eta$ ranges~\cite{Landafshitz}. Such a
classification has proved useful, both theoretically and numerically, to
develop  models which mimic such scales.  These models have the advantage of
being less complex than the original system and hence, more tractable. Indeed, we
owe much of our understanding of the physics and mathematics of turbulent
flows, validated by experiments, observations and detailed simulations, to such
reduced models. The most celebrated example of this is the tremendous advance
made within the framework of three-dimensional, homogeneous, isotropic
turbulence (in the limit of vanishing kinematic viscosity $\nu$).  Such a
framework, which ignores the specific details of the forcing  and dissipation
mechanisms, has yielded several important and universal
results~\cite{frischbook} for the inertial scale and forms the basis of
Kolmogorov's seminal work in 1941~\cite{k41}. In particular, the most important
results stemming from such a model are those related to 2-point correlation
functions, multiscaling, and their universality~\cite{frischbook,RMP,panditreview}. 

Despite the success in understanding the physics of the inertial range, the
theoretical underpinnings of the model of homogeneous, isotropic turbulence are
largely irrelevant for questions related to the regularity of such flows. This
is because these questions -- which still rank amongst the most profound and
fundamental in mathematics~\cite{clay,ee250} -- have answers hidden in the
behaviour of flows at scales much smaller than the inertial range. The key to
such answers lie in the study of incompressible, three-dimensional
Navier-Stokes and Euler equations whose non-linearity
encodes information of the velocity field at all scales ranging from the
largest to the smallest.  Remarkably, the issue of the  regularity of solutions
(for sufficiently smooth initial conditions) is still not
settled~\cite{majda-bertozzi,fmb03,bardostiti,constantin07} even for the
viscous Navier-Stokes
equation~\cite{lions,constantinfoias,temam,sohr}.  What is reasonably
clear, though, is that for initial conditions which are analytic, periodic
functions, solutions to these equations, while remaining analytic, have complex
singularities as seen, e.g., in the exponential tail of the Fourier transform
of the velocity field. In lower dimensions, such as the one-dimensional Burgers
equation with the usual (Laplacian) dissipation, the problem of finite-time
blow-up and its relation to complex singularities is completely understood~\cite{BecKhanin}. 

Although the Navier-Stokes equation, with $\nu \to 0$
and suitable initial conditions, provides a complete description of the
velocity field ${\bf u}$ in space and time, the pitfalls of a theoretical
treatment of such an equation, at small scales in particular, is best
illustrated by the following: At small scales, the properties of the solutions of 
the Navier-Stokes equation 
can be conveniently studied by neglecting the nonlinear convection term
yielding ${\hat u}_k \sim e^{ - (k/k_{\mathrm{d} })^2 }$, where $k_{\mathrm{d}
} $ is the energy dissipation wavenumber.  However,  more refined theoretical
arguments, based on the analytic properties of velocity fields at small scales
\cite{vonNeumann} or on some estimates of velocity field correlations
\cite{FMS} suggest an exponential decay as $k \to \infty $. Indeed, direct
numerical simulations suggest that the energy spectrum, at large $k$, is
consistent with the functional form $ (k/k_{\mathrm{d} } )^{\gamma } e^{-
\delta (k/k_{\mathrm{d} } ) }$ \cite{CDHKOS,MCDKWZ}. The constant $\delta $ is
believed to be a Reynolds number dependent quantity, whereas $\gamma $ is
expected to be universal. The exact numerical value of $\gamma $ is unknown and
the only prediction so far is based on Kraichnan's DIA equations
\cite{Kraichnan} giving $\gamma = 3$.

The large-$k$ asymptotic discussed above is relevant, of course, to the deep
dissipation range and have their roots in issues of regularity and finite-time
blow-up of solutions of the incompressible Navier-Stokes and Euler equations.
For moderate values of $k$, in the so-called inertial range where the ideas of
Kolmogorov hold, the energy spectrum $E(k) \sim  k^{-5/3}$ (up to intermittency
corrections). Between this intermediate and the large-$k$ asymptotics, lies the
bottleneck region. The bottleneck is defined as a bump in the turbulence
spectrum which leads to a non-monotonic behaviour in a narrow range of scales
between the inertial and dissipation range (an instance of a bottleneck in a
numerical simulation can be found in \cite{KIYIU}). It has been argued that
this pile-up is due to suppression of energy transfer to smaller scales by the
action of dissipation \cite{Falkovichbottle}. However, to the best of our
knowledge analytical predictions of the flow at bottleneck scales
have not been checked against numerical or experimental data so far, particularly  
because the pile-up effect is only weakly pronounced at the currently available Reynolds numbers.

We propose a novel approach to understand the dissipation and
the transitional (bottleneck) ranges through modifying the standard
Laplacian dissipation term $\nu \Delta {\bf u}$ by a more general function
$f(\sqrt{ - \Delta })$ (or in the Fourier space $f( \vert {\bf k} \vert )$). An
instance of such a dissipation term is the well-known hyperviscous dissipation
term $ \nu ( - \Delta )^{\alpha}$ which is frequently used in numerical
simulations. Clearly, the constants $\gamma $ and $\delta $ determining the
behavior of the energy spectrum in the dissipation range change with $\alpha $,
in particular the fall-off of the spectrum becomes steeper with growing
$\alpha$. The bottleneck has been shown to become more pronounced with
increasing $\alpha$, see, e.g., Ref.~\cite{LCP,frisch08,SPF,banerjee14} as well as 
Ref.~\cite{ssr-review} for a review, allowing
for theoretical calculations, in the large $\alpha$ limit to be checked against
numerical simulations. It is also important to remember that the use of
hyperviscosity has shed light on the problem of finite-time blow-up: It was
shown~\cite{lions,ladyzhenskaya} that there is no finite-time blow-up for
$\alpha >5/4$ despite the existence of complex singularities. For more general 
dissipative functions, there is one example that we know of where the dissipative 
term of the form $f( \vert {\bf k} \vert ) = \exp(\vert {\bf k} \vert )$ leads to entire solutions~\cite{Bardosetal}.

In this paper, by using a generalised dissipative term $f(\sqrt{ - \Delta })$,
we revisit the problem of the nature of the velocity field in the far
dissipation range as well as derive analytical results in the bottleneck region
which connects the intermediate asymptotics of the inertial range to the true
asymptotics of the far dissipation range. However, the use of such a
generalised dissipation is not completely amenable to a rigorous theoretical
treatment. This is because, as is well-known, since Euler's discovery of the
equations for ideal fluids more than 250 years ago, we are still far from
having a complete analytical handle of the nature of the velocity field in
viscous and idealised fluids. We therefore resort to a simpler model, namely that of the 
one-dimensional Burgers equation~\cite{Burgers}, which, while retaining the same structure of the 
non-linearity in the Euler and Navier-Stokes equations, allow for a more rigorous 
analytical treatment~\cite{BecKhanin,BFAuxHouches,Falkovich}. Most of our results are easily 
generalisable to higher dimensional equations of hydrodynamics; we shall comment on these 
later.

Our paper is organised as follows. 
We begin our investigations in Section~\ref{s:smallReynolds} by considering 
solutions of the one-dimensional (compressible) Burgers equation with modified dissipation 
\begin{equation}
\label{e:Burgers1}
\partial _t u + u \partial _x u =  - f( \sqrt{-\partial _x^2  } ) \, u.
\end{equation}
The approach that we use can be easily generalized to other hydrodynamical equations such as the Navier--Stokes equations. 
We show that  the leading order contribution to such solutions can be calculated recursively from an 
algebraic recursion relation involving iterated integrations. Furthermore, in the limit of large times this 
recursion relation can 
be transformed to a simple algebraic recursion relation. All of these considerations apply not only 
to hydrodynamical equation with the standard viscous term but also allow for more general 
viscous terms such as hyperviscous or exponentially growing dissipation terms~\cite{Bardosetal}. 

In Section~\ref{s:bottlesection} we investigate
the transition region between the dissipation range and the inertial range. In the case of the 
Burgers equation, the bottleneck in the spectrum is present only in the 
hyperviscous case when $\alpha > 1$. Motivated by the recent work of Frisch, {\it et al.}, \cite{SPF} 
we investigate how the presence of the bottleneck is related to Gibbs-type oscillations 
in the velocity field arising in the neighbourhood of strongly dissipating structures. In the 
framework of the one-dimensional Burgers equation the method of matched asymptotics  
can be used to derive a simplified equation for such structures which in this case 
are shocks \cite{Boyd94}. It is known that one can determine the asymptotics of the (oscillatory) solutions of 
this simplified equation. However, since neither the amplitude nor 
the phase of these oscillations is known, we determine them numerically and show that 
the asymptotic solution indeed gives the right description of the bottleneck. 
We also derive analytical relations which allow us to 
estimate for what kind of dissipation term the simplified boundary layer Burgers equation 
will exhibit a bottleneck. 

In Section~\ref{s:fullsolutions} we compare the analytical and semi-analytical
results of Sections~\ref{s:smallReynolds} and \ref{s:bottlesection} with
state-of-the-art direct numerical simulations.  By using high-precision
simulations and asymptotic extrapolation of sequences \cite{Hoeven09} we
determine the asymptotic structure of solutions in the dissipation range and
compare it with theoretical results. We also show that asymptotic solutions of
the boundary layer Burgers equation obtained in Section~\ref{s:bottlesection}
are in agreement with the numerical solutions in the bottleneck region.

%\textquestiondown 
%Finally we study the structure functions calculated for the one-dimensional
%Burgers equation by looking systematically at their small and large scale
%expansions. In particular, we study how these expansions are affected by
%hyperviscosity? 

In the last section, we discuss the implications of the results 
proved in the earlier sections and make concluding remarks. 

\section{Solutions of the Burgers equation with modified dissipation}
\label{s:smallReynolds}

The solution of the $d$-dimensional Burgers equation with standard (Laplacian)
dissipation, in the limit of vanishing viscosity, has been studied extensively,
and successfully, by using various techniques~\cite{BecKhanin,Boyd94}.
However, these established analytical approaches are limited in scope when
applied to the present problem of the Burgers equation with a dissipation term
which is not necessarily a Laplacian. This is because in the dissipation range,
such methods rarely allow us to determine beyond the leading order asymptotics.
A second drawback -- which holds even for the usual Laplacian dissipation -- is
the reliance of conventional techniques on properties peculiar to the
Burgers equation. Consequently, generalising for the higher
dimensional Euler and Navier-Stokes equations have proved formidable. 
 
Given this, we present an approach which does not rely on the specific
properties of the one-dimensional Burgers equation and hence can be generalized
to the multi-dimensional Navier--Stokes equation. This approach is in the
spirit of recent studies by Lee and Sinai of several hydrodynamic equations
with complex-valued initial conditions as well as for the case of a bounded
domain with periodic boundary conditions. Of course, we should note, that if
one were to be interested only in the one-dimensional Burgers equation, other
theoretical methods are more efficient; the advantage of our approach lies in
it being easily adapted to multidimensional equations of hydrodynamics.

We will consider two cases: (i) initial conditions which are real-valued in the
physical space but have compact support in the Fourier space and (ii) initial conditions whose
initial Fourier modes are supported on the positive half-axis (such initial
conditions are necessarily complex-valued in the physical space).

\subsection{Real-Valued Initial Conditions}
\label{s:smallReynoldslimit}

By using the semi-group $e^{-tf( \sqrt{-\partial _x^2  } ) } $, generated by $-f( \sqrt{-\partial _x^2  } )$, 
and the Duhamel principle~\cite{Yosida}, Eq.~(\ref{e:Burgers1}) can be written as
\begin{equation}
\label{e:Burgers2}
u(x, t) = - \int_0^t  e^{-(t-s) f( \sqrt{-\partial _x^2  } ) } u(x, s) \partial _x u(x, s) ds ,
\end{equation}
or in the Fourier space representation
\begin{equation}
\label{e:Burgers3}
\hat{u} (k,t) = \hat{u} (k,0) e^{-f(k) t } 
- \frac{i k}{2} 
\sum_{l+l^{\prime  } = k } e^{- f(k) t} \int_0^t 
e^{f(k) s}  \hat{u} (l,s) \, \hat{u} (l^{\prime } ,s ) \, ds.
\end{equation}
Here, and in the following, we assume that the generalised dissipation 
function $f(k)$ is a positive, non-decreasing, strictly convex even function of $k$ with 
$f(0) = 0$.

We now introduce an explicit dependence on the amplitude of the initial condition via 
\begin{equation}
\label{e:initialCondition1}
u(x, t) \vert_{t=0} = A u_0 (x) .
\end{equation}
Here, when all other parameters are fixed, $A$ plays the role of the Reynolds number.
We now expand the solution corresponding to the initial condition 
(\ref{e:initialCondition1}) in a formal power series in $A$
\begin{equation}
u(x, t) = \sum_{n=1}^{\infty } u^{(n)} (x, t) A^n ;
\end{equation}
for suitable initial conditions, this series has 
a non-vanishing radius of convergence~\cite{pauls_in_prep}. Furthermore, $u^{(n)}$ satisfy the recurrence relations
\begin{equation}
u^{(1)} = \exp \left[ - t f( \sqrt{-\partial _x^2  } ) \right] u_0 ,
\end{equation}
for $n=1$ and 
\begin{equation}
u^{(n)} = - \int_0^t \exp \left[ - (t-s) f( \sqrt{-\partial _x^2  } ) \right]
\sum_{m=1}^{n-1} u^{(m)} (s) \partial _x u^{(n-m)} (s) \, ds ,
\end{equation}
for $n > 1$.  In the Fourier space representation we use, for 
convenience, $\hat{u} (k,t) = i \hat{v} (k,t) $, so that the respective recursion relations become
\begin{equation}
\hat{v}^{(1)} (k,t)  = \hat{v} _0 (k) e^{ - f(k) t }
\end{equation}
and
\begin{equation}
\hat{v} ^{(n)} (k,t) = \frac{k}{2} \sum_{l+l^{\prime } = k } e^{ - f(k) t } 
\int_0^t e^{ f(k) s } \sum_{m=1}^{n-1}  \hat{v} ^{(m)} (l,s)
\hat{v} ^{(n-m)} (l^{\prime } , s ) \, ds.
\end{equation}
%
%It is convenient to separate the solutions for $ k > 0 $ and $ k < 0 $, respectively. Hence for $k>0$, 
%\begin{eqnarray}
%\hat{v} ^{(n)} (k,t) &=&  \frac{k}{2} \sum_{l=1}^{k-1}  e^{ - f(k) t } \int_0^t e^{ f(k) s } 
%\sum_{m=1}^{n-1} \hat{v} ^{(m)} ( l , s ) \hat{v} ^{(n-m)} ( k - l , s ) \, ds 
%\nonumber \\ 
%&+& 
%k \sum_{l=1}^{ \infty }  e^{ - f(k) t } \int_0^t e^{ f(k) s } 
%\sum_{m=1}^{n-1} \hat{v} ^{(m)} ( k + l , s ) \hat{v} ^{(n-m)} ( - l , s ) \, ds; 
%\end{eqnarray}
%an analogous formula can be written for $ k < 0 $. Now we make the following observation:
%
By using these recursive formulas it is easy to make the following observation:
\begin{Prop}
Suppose that $ \hat{v} _0 (k) $ is supported only by finitely many modes 
in the Fourier space, i.e., $ k \in \Sigma = \{ - K , ... , K \} $, where $K$ is a 
positive integer.
Then for any $ n \geq 2 $ the function 
$ \hat{v} ^{(n)} (k,t) $ also has a finite support in the Fourier space
$ n \Sigma = \{ - n K , ... , n K \} $ and every 
$\hat{v} ^{(n)} (k,t) $ can be calculated by finitely many operations. 
\end{Prop}

This observation can be used to calculate recursively solutions of Eq.(~\ref{e:Burgers1}) 
in a manner similar to the recursive calculation of 
solutions of inviscid equations by means of time series expansions. 
Here we demonstrate this on the example of initial conditions consisting of 
one Fourier mode $u_0 (x) = \sin x$. Then the first two terms of the expansion are
\begin{equation}
u^{(1)} (x, t) = \sin x \, e^{-tf(1)} ,
\end{equation}
and 
\begin{equation}
u^{(2)} (x, t) = - \frac{1}{2} \frac{1}{f(2) - 2f(1)} \, \sin 2x \, 
\Bigl( e^{-2f(1)t} - e^{-f(2)t} \Bigr) . \nonumber 
\end{equation}
It is not difficult to verify that the higher order terms are of the form
\begin{equation}
u^{(n)} (x, t) = g^{(n)}_0 (t) \, \sin n x + g^{(n)}_2 (t) \, \sin (n-2) x + ... \ .
\end{equation}
This representation can be easily transferred into the Fourier representation as
\begin{eqnarray}
v^{(n)} (n, t)   = - \frac{1}{2} \, g_0^{(n)} (t) , \qquad
&
 v^{(n)} (n-2, t) = - \frac{1}{2} \, g_2^{(n)} (t) 
&
, \qquad ... 
\\   
v^{(n)} (-n, t)   = \frac{1}{2} \, g_0^{(n)} (t) , \qquad
& 
v^{(n)} (-n+2, t) = \frac{1}{2} \, g_2^{(n)} (t) 
&
, \qquad ...   
\end{eqnarray}
We now note that although for a fixed wavenumber $k$ obtaining $ \hat{v} (k,t) $ 
requires summation of the whole series in $A$, to obtain small-$A$ asymptotics 
it suffices to consider only the lowest power of $ A $ for a given $k$. For the initial
condition $\sin x $ this lowest power for wavenumber $k$ is $A^k$ and we obtain the following
small $A$ asymptotics
\begin{eqnarray}
\hat{v} (k,t) &\sim& \hat{v} _{\mathrm{as} }  (k,t) = A^{-k} \hat{v} ^{(-k)} (k,t) \quad k < 0; 
\\
\hat{v} (k,t) &\sim& \hat{v} _{\mathrm{as} }  (k,t) = A^k \hat{v} ^{(k)} (k,t) \quad  k > 0.
\end{eqnarray}
The function $ \hat{v} _{\mathrm{as} }  (k,t) $ can be represented as a sum of two functions 
$ \hat{v} _{\mathrm{as} }^{+}  (k,t) $ and $ \hat{v} _{\mathrm{as} }^{-}  (k,t) $ with support
on the positive and negative half-axis, respectively. The function 
$ \hat{v} _{\mathrm{as} }^{+}  (k,t) $ is the  solution of the recurrent equation (for $k>0$) 
\begin{equation}
\label{e:Burgerscomplex1}
\hspace*{-1.8cm}
\hat{v} _{\mathrm{as} } ^{+} (k,t) = \hat{v} (k,0) e^{-f(k) t }  +  \frac{k}{2} 
\sum_{l = 1 }^{k-1}  e^{- f(k) t} \int_0^t 
e^{f(k) s}  \hat{v} _{\mathrm{as} }^{+} (l,s) \, \hat{v} _{\mathrm{as} }^{+} (k - l , s ) \, ds.
\end{equation}
The function $ \hat{v} _{\mathrm{as} } ^{-} (k,t) $ satisfies an analogous equation with $k < 0$. 
Let us stress that the functions  $ \hat{v} _{\mathrm{as} } ^{+} (k,t) $ and 
$ \hat{v} _{\mathrm{as} } ^{-} (k,t) $ are also solutions of the Burgers equation, however, 
with one-mode {\it complex-valued} initial conditions. The function $ \hat{v} _{\mathrm{as} } ^{+} (k,t) $
corresponds to the initial condition $ \hat{v}  _{\mathrm{as} } ^{+} (1,0) = A \hat{v} _0 (1)  $ 
and the function $ \hat{v} _{\mathrm{as} } ^{-} (k,t) $ to the initial conditions 
$ \hat{v}  _{\mathrm{as} } ^{-} (-1, 0) = A \hat{v} _0 (-1)  $. The asymptotic solution 
in the physical space $ u_{\mathrm{as} }  (x,t) $
can be written in terms of the coefficients $g_0^{(n)} (t)$ 
\begin{equation}
u_{\mathrm{as} }  (x,t) = \sum_{n=1}^{\infty } g_0^{(n)} (t) \, \sin nx \, A^n ,
\end{equation}
where
\begin{equation}
\hspace*{-1.8cm}
g^{(n)}_0 (t)  = 
\sum_{\alpha _1 + 2 \alpha _2 ...  + n \alpha _n = n} \frac{1}{2^{n-1} } \, 
G(n; \alpha _1 , \alpha _2 ,..., \alpha_n ) \, 
e^{ -t (\alpha _1 f(1) + \alpha _2 f(2) + ... + \alpha _n f(n) ) } .
\label{gn}
\end{equation}
The sum is taken over all combinations of $n$ non-negative integers $\alpha _i $, 
$i = 1, ..., n$ such that $ \alpha _1 + 2 \alpha _2 + ... + n \alpha _n = n$. We note that 
these combinations correspond to partitions of integers.
For example, for $n$ from $1$ to $4$, we obtain the following combinations:
\begin{itemize}
\item
$n = 1$: $\{(\alpha _1 ) \} = \{ (1) \} $,
\item
$n = 2$: $\{(\alpha _1 , \alpha _2 ) \} = \{ (2,0) , (0, 1) \} $,
\item
$n = 3$: $\{(\alpha _1 , \alpha _2 , \alpha _3 ) \} = \{ (3, 0, 0) , (1, 1, 0) , (0, 0, 1) \} $,
\item
$n = 4$: $\{(\alpha _1 , \alpha _2 , \alpha _3 , \alpha _4 ) \} = \{ (4, 0, 0, 0) , (2, 1, 0, 0) , (0, 2, 0, 0) ,
(1, 0, 1, 0) , (0, 0, 0, 1) \} $.
\end{itemize}
The recurrence relation for coefficients $G(n;(n,0,...,0))$, $n\geq 1$ corresponding 
to combinations $(n, 0, ..., 0)$ is 
$$
G(n;(n, 0, ..., 0)) = - \frac{n}{2} \, \frac{1}{f(n) - n f(1)} \, \sum_{m=1}^{n-1}  G(m; (m, 0, ..., 0) \, G(n-m; (n-m, 0, ..., 0)).
$$
Note that these coefficients contribute to terms with decay rate $e^{-ntf(1)} $, which is
the slowest decay rate possible for $g_0^{(n)} (t)$. 
For coefficients $G(n;(n-2, 1, 0, ..., 0))$, $n \geq 3$ of combinations 
$(n-2, 1, 0, ..., 0)$, we obtain
\begin{eqnarray}
\hspace*{-1.8cm}
G(n;(n-2, 1, 0, ..., 0)) 
= - \frac{n}{f(n) - (n-2) f(1) - f(2)} \, 
G(n-1, (n-3, 1, 0, ..., 0)) - 
\nonumber
\\
\hspace*{-1.8cm}
\frac{n}{2} \, \frac{1}{f(n) - (n-2) f(1) - f(2)} \, \sum_{m=2}^{n-2}  
\biggl[ G(m; (m, 0, ..., 0) \, G(n-m; (n-m-2, 1, 0, ..., 0)) + 
\nonumber
\\
\hspace*{-1.8cm}
G(m; (m-2, 1, 0, ..., 0) \, G(n-m; (n-m, 0, ..., 0)) \biggr] .
\end{eqnarray}
Generally, for coefficients  
$G(n; (\alpha _1 , ..., \alpha _{n-1} , 0))$ of combinations of the type
$(\alpha _1 , ..., \alpha _{n-1} , 0)$, where the last entry vanishes, we obtain the relation
\begin{eqnarray}
G(n; (\alpha _1 , ..., \alpha _{n-1} , 0)) = - \frac{n}{2} \, 
\frac{1}{f(n) - \alpha _1 f(1) - ... - \alpha _{n-1} f(n-1)} 
\nonumber
\\ 
\sum_{m=1}^{n-1} \sum_{ \underline{\beta } + \underline{\gamma } = \underline{\alpha }}
G(m; (\beta _1 , ..., \beta _m ) \, G(n-m; (\gamma _1 , ..., \gamma _{n-m} )) .
\end{eqnarray}
Here we denote $\underline{\alpha } = (\alpha _1 , ..., \alpha _{n-1} ) \in \mathbb{N} _0^{n-1}$,
$\underline{\beta } = (\beta _1 , ..., \beta m , 0, ..., 0) \in \mathbb{N} _0^{n-1}$ and
$\underline{\gamma } = (\gamma _1 , ..., \gamma _{n-m} , 0, ..., 0) \in \mathbb{N} _0^{n-1}$, 
where $\beta _1 + 2 \beta _2 + ... + m \beta _m = m$ and 
$\gamma _1 + 2 \gamma _2 + ... + (n-m) \gamma _{n-m} = n - m$.
Coefficients $G(n; (0,0, ..., 0,1))$ can be calculated from the coefficients corresponding 
to combinations with $\alpha _n = 0 $
\begin{eqnarray}
\hspace*{-2.4cm}
G(n; (0,0, ..., 0,1)) =  \sum_{m=1}^{n-1} (n-m) 
\sum_{\alpha _1 + 2 \alpha _2 ... + m \alpha _m = m} \ 
\sum_{\beta _1 + 2 \beta _2 ...  + (n-m) \beta _{n-m} = n-m} 
\nonumber
\\
\hspace*{-2.4cm}
\frac{G(m; \alpha _1 , \alpha _2 , ... ,\alpha _m ) \, G(n-m; \beta _1 , \beta _2 , ... ,\beta _{n-m} ) }{f(n) - 
\alpha _1 f(1) - \alpha _2 f(2) - ... - \alpha _m f(m) - \beta _1 f(1) - \beta _2 f(2) - ... - 
\beta _{n-m} f(n-m)}.
\end{eqnarray}
We note that in the case, when the function $f(\cdot )$ describe the standard 
dissipation $-\nu \partial _x^2 $ the coefficients $G(\cdot ; \cdot )$ can 
be found explicitly by means of the Hopf-Cole transformation which yields the familiar
expression
\begin{equation}
u (x, t) = -2\nu \ln \biggl( 
e^{ t \nu \partial _x^2 } \, e^{ - \frac{A}{2\nu } u_0 } 
\biggr) .
\end{equation}

\subsection{Solutions of the one-dimensional Burgers equation for complex-valued initial conditions}
\label{ss:complexinitialdata}

As has been noted in Ref.~\cite{Bardosetal}, for initial conditions supported on the positive 
half-line, i.e.,  
$\hat{v} ( k , 0 ) = 0 $ for $ k \leq 0 $, the Fourier coefficient of 
the solution at a fixed wavenumber can be calculated iteratively by finitely many 
operations via Eq.~(\ref{e:Burgerscomplex1}). Thus, e,g., we obtain 
\begin{equation}
\hat{v} (1,t) =  \hat{v} _0 (1) \, e^{- f (1) t } ,
\end{equation}
\begin{equation}
\hat{v} (2 , t) =  \Bigl[ \hat{v} _0 (2) - \frac{1}{ f(2) - 2 f(1) }  \hat{v} _0^2 (1) \Bigr] 
e^{- f (2) t }  + 
\frac{ \hat{v} _0^2 (1) e^{ - 2 f(1) t }}{ f(2) - 2 f(1) }
\end{equation}
and
\begin{eqnarray}
\hspace*{-1.5cm}
\hat{v} (3,t) &=& \Biggl\{ 
\hat{v} _0 (3) - \frac{ 3 \hat{v} _0 (1) \hat{v} _0 (2)}{f(3) - f(1) - f(2) } \nonumber \\ &+& 
\frac{3 \hat{v} _0^3 (1)}{f(2) - 2 f(1) } \, \biggl[ \frac{1}{f(3) - f(1) - f(2)} - 
\frac{1}{f(3) - 3 f(1)} \biggr] \Biggr\} \, e^{ - f(3) t }   
\nonumber \\  &+& 
3 \, \biggl[ \hat{v} _0 (1) \hat{v} _0 (2) - \frac{ \hat{v} _0^3 (1)}{f(2) - 2f(1)} \biggr]
\frac{ e^{ - [ f(1) + f(2) ] t}}{f(3) - f(1) - f(2)} \nonumber \\ &+& 
 \frac{3 \hat{v} _0^3 (1)  e^{ - 3 f(1) t }}{[f(2) - 2f(1)][f(3) - 3f(1)]}.
\end{eqnarray}
In general the Fourier coefficients of the solution will have a form which is similar to 
Eq.~(\ref{gn})
\begin{equation}
\hspace*{-1.8cm}
\label{e:generalsol}
\hat{v} (k,t) = \sum_{\alpha _1  + 2 \alpha _2 + ... + k \alpha _k = k } 
F (k; (\alpha _1, \alpha _2 , ..., \alpha _k ) ) \, 
e^{-t(\alpha _1 f(1) + \alpha _2 f(2) + ... \alpha _k f(k) )} . 
\end{equation}
It is instructive to compare this form with the explicit expression found in the case 
of $f( \sqrt{-\partial _x^2  } ) = - \nu \partial _x^2 $ and $u_0 (x) = A e^{ix} $
obtained by using Fa\`a di Bruno's formula
\begin{eqnarray}
\hat{u} (k, t) = -2\nu \, \frac{(-1)^k }{k!} \frac{A^k }{2^k \nu ^k } \,
\sum_{l=1}^k (-1)^{l-1} (l-1)! \,
\sum_{j_1 , j_2 , ..., j_{k-l+1}} \frac{k!}{j_1 ! j_2 ! ... j_{k-l+1} !} \times
\nonumber
\\
\left( \frac{1}{1!} \right)^{j_1 }  \left( \frac{1}{2!} \right)^{j_2 }  ...
\left( \frac{1}{(k-l+1)!} \right)^{j_{k-l+1} } \,
e^{-\nu t (j_1 + j_2 2^2 + ... j_{k-l+1} (k-l+1)^2 )} ,
\end{eqnarray}
where the second sum is taken over $k - l + 1$ nonnegative integers $j_1 , ..., j_{k-l+1}$ such that
\begin{displaymath}
j_1 + j_2 + ... + j_{k-l+1} = l ,
\end{displaymath}
and
\begin{displaymath}
j_1 + 2 j_2 + ... + (k-l+1) j_{k-l+1} = k .
\end{displaymath}
In the explicit solution, the dependence on the amplitude of the initial condition $A$ 
manifests itself by the term $A^k$, in agreement with the observation made previously.
The time dependence is also clearly exhibited by the terms
 $e^{-\nu t (j_1 + j_2 2^2 + ... j_{k-l+1} (k-l+1)^2 )} $, which in the more general case become 
$e^{-t(\alpha _1 f(1) + \alpha _2 f(2) + ... \alpha _k f(k) )}$. Finally, the exact solution also
gives explicit expressions for the coefficients 
$F (k; (\alpha _1, \alpha _2 , ..., \alpha _k ) )$. 

In general, calculating solutions of the Burgers equation with modified dissipation
by recursive determination of coefficients $F (k; (\alpha _1, \alpha _2 , ..., \alpha _k ) )$ 
is quite cumbersome. We now take advantage of the fact that in the limit of large times 
the terms with exponential decay $e^{-kf(1)} $ dominate over the other terms.
\begin{Prop}
From the assumptions on $f(\cdot ) $, it follows that $f(\cdot )$ is a super-additive function. 
The term in the sum on the right-hand-side of Eq.~(\ref{e:generalsol})
with the slowest decay in $t$ corresponds to 
$ (\alpha _1 , \alpha _2 , ... , \alpha _k ) = (k, 0, ..., 0) $, with the
rate of temporal decay $e^{ - k f(1) } $. The coefficient 
$F ( k, (k,0,...,0) ) = h(k) $  satisfies the following recursion relation
\begin{equation}
\label{e:recursion1}
h (k)  = \frac{k}{2} \frac{1}{f(k) - k f(1) } \sum_{l=1}^{k-1} 
h ( l ) \, h  ( k -l )  ,
\end{equation}
with $h ( 1) = \hat{v} _0 (1 ) $. Thus, for a fixed $k$ and $t \to + \infty $
\begin{equation}
\hat{v} (k,t) \sim F ( k , (k, 0, ... , 0) ) \, e^{ - k f(1) t } .
\end{equation}
\end{Prop}
Note that the high wavenumber contributions to the initial conditions are suppressed
when $t \to \infty $ and the solution therefore becomes independent of the initial 
condition since the $k$th mode is proportional to $\hat{v} _0^k (1) $. 
Thus, the behaviour of solutions of Eq.~(\ref{e:Burgers3}) is universal at large 
times. 

\subsection{Reduction to an ordinary difference-differential equation}
\label{s:difference-differential}

To study the solutions of the recursion relation Eq.~(\ref{e:recursion1}), we introduce
the generating function $h(x)$ of $ h(k)$
\begin{equation}
h(x) = \sum_{k=1}^{\infty }  h(k) \, e^{kx} ,
\end{equation}
so that Eq.~(\ref{e:recursion1}) becomes an ordinary pseudo-differential equation
\begin{equation}
\label{e:Burgersfinal}
f  \left( \partial _x \right ) h  - f(1) \partial _x h = \frac{1}{2}  \partial _x h^2 , 
\end{equation}
with boundary conditions $  h (x)  \sim  \hat{v} _0 (1) e^{x} $ for $x \to - \infty $.
It is well-known that the asymptotic properties of $h(k)$ can be deduced from the 
analytic properties of $ h(\xi ) $. Here we will consider two cases: (i) hyperviscosity 
\begin{equation}
f_1 ( k ) = k^{ 2 \alpha } ,
\end{equation}
and (ii) exponentially growing dissipation
\begin{equation}
f_2 ( k ) = e^k . 
\end{equation}

In case (i) the solution $ h (x) $ has a singularity at some point $ x_0 $. 
We know that the solution of Eq.~(\ref{e:Burgersfinal}) in the neighborhood of the 
singularity behaves as 
$ ( x - x_0 )^{ 2 \alpha - 1 } $
\begin{equation}
\label{e:hyperviscous1}
h(x) = \frac{1}{ ( x - x_0 )^{2 \alpha - 1} } \, g ( x - x_0 ) .
\end{equation}
To determine higher-order contributions, we assume 
that the function $ g ( x - x_0 ) $ can be written as
\begin{equation}
\label{e:hyperviscous2}
g ( x - x_0 ) = g^{(1)} ( x - x_0 ) + ( x - x_0 )^{\gamma } g^{(2)} ( x - x_0 ) + {\rm h.o.t.},
\end{equation}
where the functions $ g^{(1)} ( x - x_0 ) $ and $ g^{(2)} ( x - x_0 ) $ are analytic
with Taylor expansions
\begin{equation}
g^{(1)}  ( \xi ) = \sum_{l=0}^{\infty } g_l^{(1)} \xi ^l , \quad  
g^{(2)}  ( \xi ) = \sum_{l=0}^{\infty } g_l^{(2)} \xi ^l ;
\end{equation}
and the remaining terms are of higher, non-integer orders. Inserting 
the representations (\ref{e:hyperviscous1}) and (\ref{e:hyperviscous2})
into Eq.~(\ref{e:Burgersfinal}), we obtain the following equation for $\gamma $
\begin{equation}
\sum_{m=0}^{2 \alpha - 2} (-1)^m {{2\alpha - 1}\choose{m}}
\frac{ ( 2 \alpha - 2 + m)! }{ (2 \alpha - 2)! } ( \gamma )_{2 \alpha - 1 - m } + 
\frac{ (4 \alpha - 3)! }{ (2 \alpha - 2)! } = 0 .
\end{equation}
Here $ ( \gamma )_{ 2 \alpha - m - 1} $ is the Pochhammer symbol. 
One solution is $ \gamma = - 1$; the other solutions are complex and we 
denote them by $ \gamma _i^{\pm } $, $ i = 1,...,\alpha - 1 $, with 
$ \mathit{Re} ( \gamma _i^{\pm } ) > 0 $ and $ ( \gamma _i^{+} )^{\ast } = 
 \gamma _i^{-} $. The terms $ (x - x_0 )^{ \gamma _i^{\pm }  } $ imply that the 
asymptotic expansion of  $ h(k)$  for $k \to \infty $ has the form
\begin{equation}
\label{e:hypervisous1}
h(k) \simeq C k^{ 2 \alpha - 2 } e^{ - \delta k } \Bigl( 
1 +  \frac{b_1 }{k } + ... + \sum_{i=1}^{\alpha - 1} 
c_1^{i} k^{ - \gamma _i^{+} } + \sum_{i=1}^{\alpha - 1} 
( c_1^{i} )^{\ast } k^{ - \gamma _i^{-} } + ... 
\Bigr) 
\end{equation}

For case (ii), Eq.~(\ref{e:Burgersfinal}) becomes a difference-differential equation
\begin{equation}
h(x+1) = \partial _x \Bigl\{ \frac{1}{2} h^2 (x) + e^1 h(x) \Bigr\} .
\end{equation}
Solutions of this equation are entire functions~\cite{Bardosetal}; therefore we concentrate on their behaviour 
for $ x \to \infty $. Assuming that $h(x) \to \infty $ for $ x \to \infty $ we write $ h(x) = \exp [ S(x) ] $ 
obtaining
\begin{equation}
e^{ S(x+1) } = \partial _x \Bigl\{ e^{ 2 S(x) } + e^{ S(x) +1 } \Bigr\}  .
\end{equation}
The dominant behaviour can be deduced from the relation
\begin{equation}
S (x+1) = 2 S (x) ,
\end{equation}
which is solved by
\begin{equation}
S(x) = \beta (x) \, e^{ x \ln 2 }  .
\end{equation}
Here $ \beta (x) $ is a periodic function with period $1$: 
$ \beta (x+1) = \beta (x) $. Thus, to the leading order the solution is given by
\begin{equation}
H(x) = \exp \left[  \beta (x) e^{x \ln 2 }  \right] .
\end{equation}
From this representation it is easy to determine the behaviour of $ h(k)$ as 
follows: Introducing a new variable $ \xi = e^x $ we see that $h(k)$ are 
the Taylor coefficients of $ \tilde{h} (\xi ) = h ( \ln e^x ) $ and that for $ \xi \to \infty $ 
\begin{equation}
\tilde{h} (\xi ) \sim \exp \left[ \beta ( \ln \xi ) \, \xi ^{\ln 2} \right] .
\end{equation}
The function $\tilde{h} (\xi ) $ is thus an entire function of order $ \ln 2 $. 
It is well-known that the growth rate of entire functions at infinity determines
the behavior of their Taylor coefficients for $ k \to \infty $~\cite{Levin} so that
\begin{equation}
\label{e:ExpDissAsEstimation}
h(k) \sim e^{ - \frac{1}{\ln 2 } k \ln k } .
\end{equation}
Actually, the asymptotic behaviour of $ h(k)$ can be determined directly from
the recursion relation for $h(k)$ along with sub-dominant terms
\begin{equation}
\label{e:cosh-dissipation1}
h(k) \simeq \frac{1}{2^{\frac{3}{2} } \sqrt{ \pi  \ln 2 } } k^{ - \frac{3}{2} } 
e^{ (\delta + g ( \ln k ) ) k } \, e^{ - \frac{1}{\ln 2 } k \ln k } ,
\end{equation}
where the function $ g(\cdot )$ is periodic with period $ \ln 2 $.
The presence of the function $ g( \ln k ) $ in the asymptotic expansion of 
$h(k)$ is related to the presence of the function $\beta ( x ) $ in the expansion 
of $ h(x)$ at infinity. 

Finally, we remark that the estimate~(\ref{e:ExpDissAsEstimation})
of the dominant part in the high wavenumber 
asymptotics of solutions of the Burgers equation with exponentially growing dissipation
can be proved rigorously \cite{pauls_in_prep}.

\section{Bottleneck effect in the boundary layer of the one-dimensional Burgers equation}
\label{s:bottlesection}

The analysis presented in the previous section applies only to 
small Reynolds numbers and can thus be relevant only for the dissipation 
range. To study the transition zone between the dissipation range and
the inertial range we have to take recourse to asymptotic matching which so
far is known to work only for the Burgers equation.
We write the Burgers equation with modified dissipation in the form
\begin{equation}
\label{e:Burgers11}
\partial _t u + u \partial _x u = - \frac{1}{\nu } f( \nu \sqrt{-\partial _x^2  } ) \, u .
\end{equation}
In the limit of vanishing viscosity $\nu \to 0$ the {\it outer} solution, which is the 
entropic solution of the inviscid Burgers equation, is matched against 
the {\it inner} solution of the equation
\begin{equation}
\label{e:innerBurgers1}
f  \left( \sqrt{ - \frac{d^2}{dX^2} }  \right) \, u^{(\mathrm{in} )} + u^{(\mathrm{in} )} 
\frac{d}{dX} u^{(\mathrm{in} )} = 0 , 
\end{equation}
satisfying the boundary conditions 
$ \lim_{X \to  - \infty } u^{(\mathrm{in} )}  ( X ) = 1 $ and 
$ \lim_{X \to  + \infty } u^{(\mathrm{in} )}  ( X ) = - 1 $~\cite{SPF,Boyd92}. 
In this section we shall study various aspects of solutions of the inner equation
(\ref{e:innerBurgers1}), in particular, with an eye on the bottleneck effect.

\subsection{Bottleneck and oscillations}
\label{s:bottleoscillations}

Equation~(\ref{e:innerBurgers1}), for the case of the hyperviscous dissipation term 
$ f(k)  = k^{2 \alpha } $,  has been studied by asymptotic and 
numerical methods in Refs.~\cite{SPF,Boyd94}.  The same methods are easily extended to 
Eq.~(\ref{e:innerBurgers1}) for more general dissipation terms. Thus, in the
case of a more general $f(k)$ (here, we only assume that it grows faster than linearly)
we can use asymptotic expansion of the solution at $\pm \infty $:
neglecting nonlinear contributions Eq.~(\ref{e:innerBurgers1}) 
is written as
\begin{eqnarray}
& f  \left( \sqrt{ - \frac{d^2}{dX^2} }  \right) \, u^{(\mathrm{in} )}_{\mathrm{as} } + 
\frac{d}{dX}  u^{(\mathrm{in} )}_{\mathrm{as} } = 0 , & X \to  - \infty ,
\\
& 
f  \left( \sqrt{ - \frac{d^2}{dX^2} }  \right) \, u_{\mathrm{as} }^{(\mathrm{in} )} - 
\frac{d}{dX}  u_{\mathrm{as} }^{(\mathrm{in} )} = 0 , & 
X \to  + \infty .
\end{eqnarray}
By using the {\it ans\"atz} $ u_{\mathrm{as} }^{(\mathrm{in} )} = e^{ - i \zeta x }$  and 
$ u_{\mathrm{as} }^{(\mathrm{in} )} = e^{ i \zeta x }$ we obtain an equation
for $\zeta $
\begin{equation}
\label{e:zeta}
\frac{1}{\zeta } f( \zeta ) = i .
\end{equation}
The boundary conditions imply that we take only those solutions 
$ \zeta _i $ of Eq.~(\ref{e:zeta}) for which 
$\mathrm{Im} \, \zeta _i > 0 $.  Let us consider the solution
$\zeta _{\mathrm{min} } $ of Eq.~(\ref{e:zeta}) 
with the smallest imaginary part $\mathrm{Im} \, \zeta _{\mathrm{min} }$. 
The leading order asymptotics for $ X \to \pm \infty $ are 
\begin{eqnarray}
\label{e:leading-asymptoticsBLHBurgers}
u^{(\mathrm{in} )} (X) &\simeq&  u^{(\mathrm{in} )}_{\mathrm{as} } (X) = 
1 -  A  e^{ \lambda  X }
\sin ( \omega   X + \phi )  , 
\qquad X \to - \infty; \nonumber
\\
u^{(\mathrm{in} )} (X) &\simeq& u^{(\mathrm{in} )}_{\mathrm{as} } (X) =  
- 1 - A  e^{ - \lambda   X }
\sin ( \omega  X -  \phi  ) , 
\qquad X \to + \infty ;
\end{eqnarray}
where $ \lambda = \mathrm{Im}\, \zeta $ and $\omega = \mathrm{Re} \,
\zeta $ and $A$ is chosen  to be positive.  
In the case when $ \mathrm{Re} \,
\lambda _{\mathrm{min} } \neq 0 $, 
the solution oscillates around $ \pm 1 $. However, neither the 
amplitude of the oscillation $A$, nor the phase $\phi $ can be determined from
a linear analysis. 

%Let us now consider the relation between the oscillations in  the solutions and 
%the presence of a bottleneck. 
The Fourier transform of  the linearized solution is
\begin{equation}
\label{e:leading-asymptoticsBLHBurgers1}
\hspace*{-2cm}
\hat{u} ^{(\mathrm{in} )}_{\mathrm{as} } (k) =   - i \sqrt{ \frac{2}{\pi } } \frac{1}{k}   - 
i \sqrt{\frac{2}{\pi } }  \, A \, k \, 
\frac { k^2 \sin \phi + \lambda ^2 \sin \phi  - 2\, \lambda \, \omega \, \cos  \phi  
-  \omega ^2 \, \sin  \phi }{ \left(  \lambda ^2 + \omega ^2 - 2 \, \omega \, k + k^2 
\right)  \left(  \lambda ^2 +  \omega ^2 + 2\, \omega\, k + k^2 \right) }  .
\end{equation}
At high $k$ the linearized asymptotic solution $ \hat{u} ^{(\mathrm{in} )}_{\mathrm{as} } (k) $ 
decays as $ - i \textstyle\sqrt{2/\pi } \, (1 + A \sin \phi ) \, k^{-1} $ contrary to the actual solutions of 
Eq.~(\ref{e:innerBurgers1}) which decay exponentially or faster 
than exponentially. Nevertheless, we expect that the dissipation range is being mimicked 
also for  the linearized asymptotic solution:  For high $k$ the asymptotic solution has to
decrease faster than the small $k$ solution $ - i \textstyle\sqrt{2/\pi }  \, k^{-1} $ and this is possible
only when $ A \sin \phi < 0 $ as confirmed by numerical simulations.  

\begin{figure*}[htbp]
   \centering
   \includegraphics[scale=0.6]{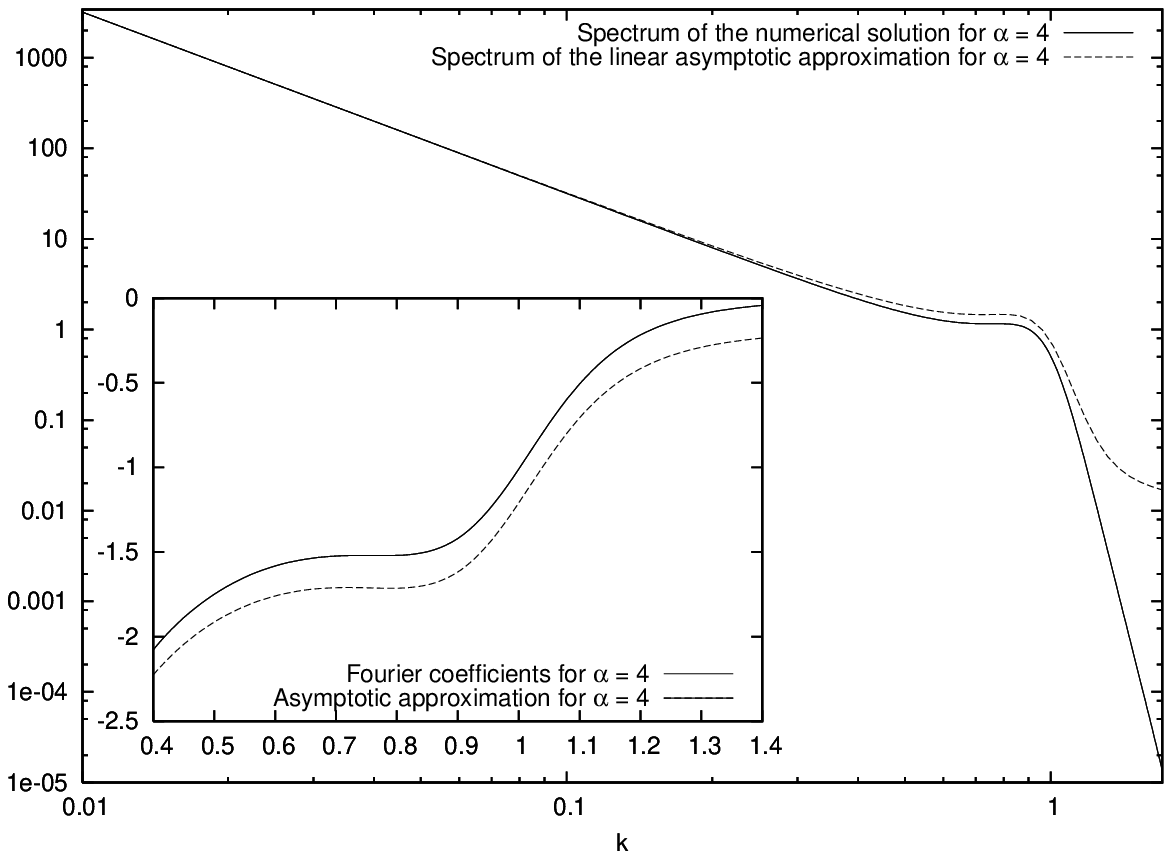} 
   \includegraphics[scale=0.6]{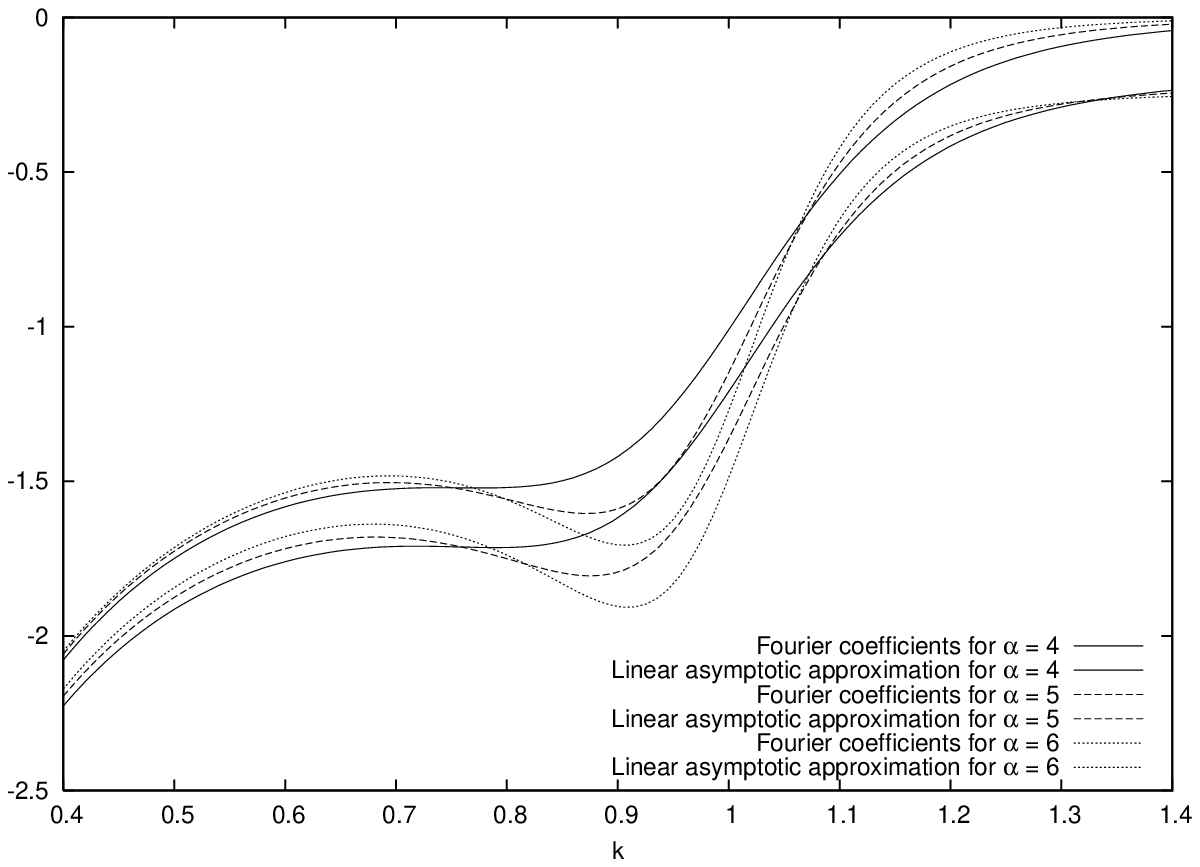} 
   \caption{
   		Comparison between the imaginary part of numerical solution of Eq.~(\ref{e:innerBurgers1})  in the Fourier space and  the 
		imaginary part of the linearized asymptotic solution Eq.~(\ref{e:leading-asymptoticsBLHBurgers})
		in the Fourier space for $ f(k) = k^8 $  (solid lines), $f(k) = k^{10} $ (dashed lines) and
		$f(k) = k^{12} $ (dotted lines). Note that in the bottleneck region the numerical and the 
		asymptotic solutions have the same shape with the asymptotic solution shifted down compared to
		the numerical solution. 
	   	}
   \label{fig:linbottleneck}
\end{figure*}

In Fig.~(\ref{fig:linbottleneck}), we compare numerical solutions and linearized asymptotic solutions for the hyperviscous 
dissipation term $f(k) = k^{2 \alpha } $
 with $ \alpha = 4, 5, 6$.
The agreement between the two is remarkably good, in particular in the bottleneck region they seem to have 
similar shapes. We do note, however, that the linear asymptotic solution is shifted with respect to the complete
solution. Thus, although the expression~(\ref{e:leading-asymptoticsBLHBurgers1}) is an excellent model 
for the solution in the bottleneck region, there is a drawback: The amplitude $A$ and the phase shift 
$\phi $ cannot be determined analytically and has to be extracted numerically.

Now we derive an integral identity for solutions of Eq.~(\ref{e:innerBurgers1}).
We multiply Eq.~(\ref{e:innerBurgers1}) by $G( u^{(\mathrm{in} ) } ) $
and integrate over $X$, obtaining
\begin{equation}
\label{e:integral1}
\int_{\mathbb{R} }  G ( u^{( \mathrm{in} ) } ) f  \left( \sqrt{ - \frac{d^2}{dX^2} }  \right) 
 u^{( \mathrm{in} ) }  \, dX = 
%\int_{\mathbb{R} }  f  (k) \vert \hat{u} ^{( \mathrm{in} ) } (k) \vert ^2 \, dk = \frac{2}{3} .
g(1) - g(-1) ,
\end{equation}
where the functions $G(\cdot ) $ and $g ( \cdot ) $ are related by
\begin{equation}
\frac{d}{du} g (u) = u G(u) .
\end{equation}
For $ G(u) = u $, we obtain the relation
\begin{equation}
\label{e:integral2}
\int_{\mathbb{R} }  f  (k) \vert \hat{u} ^{( \mathrm{in} ) } (k) \vert ^2 \, dk = \frac{2}{3} .
\end{equation}
We divide the Fourier space into three ranges: the small wave-number range
$ ( - k_{\mathrm{i} } , k_{\mathrm{i} } ) $, corresponding to the inertial range, 
the intermediate wave-number range $ ( - k_{\mathrm{d} } , - k_{\mathrm{i} } ) 
\cup ( k_{\mathrm{i} } , k_{\mathrm{d} } ) $ 
and the high wave-number range $ ( - \infty , - k_{\mathrm{d} } ) \cup 
( k_{\mathrm{d} } , + \infty  ) $ which corresponds to the dissipation range. 
Because of the exponential decay of the Fourier coefficients in the dissipation range the 
contribution to the integral~(\ref{e:integral2}) from the high 
wave-number range is negligible.

To a first approximation, we estimate the width of the small wave-number range 
$ ( - k_{\mathrm{i} } , k_{\mathrm{i} } ) $ 
by assuming that the entire contribution to~(\ref{e:integral2}) comes from this range
\begin{equation}
\label{e:bottleneckestimate1}
\frac{2}{\pi } \int_{ - k_{\mathrm{i} } }^{k_{\mathrm{i} }  }  \frac{ f(k) }{k^2 } \, dk  = \frac{2}{3} .
\end{equation}
Obviously, the solution of Eq.~(\ref{e:bottleneckestimate1}) gives an upper bound for 
the higher end of the inertial range. 
By setting $f( k_{\mathrm{d} } ) = 1 $ at the lower end of the dissipation range, we 
estimate the beginning of the dissipation range. 
For the definitions of $ k_{\mathrm{i} } $ and $ k_{\mathrm{d} } $ to be consistent we require
that $ k_{\mathrm{i} } <  k_{\mathrm{d} } $. However, for $f(k)$ such that $ f(k) / k^2 $ are
small for small $k$ this consistency condition is violated. 

\begin{figure*}[t]
   \centering
   \includegraphics[scale=0.6]{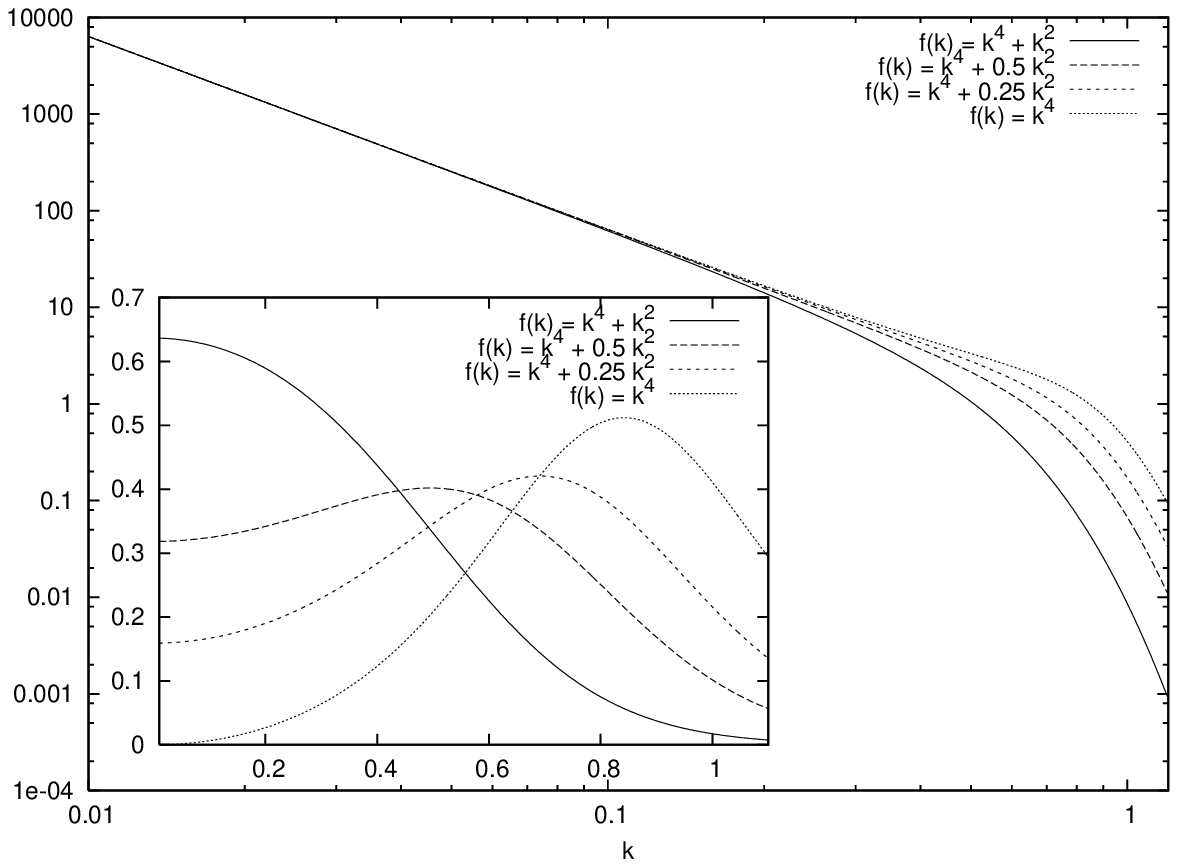} 
   \includegraphics[scale=0.6]{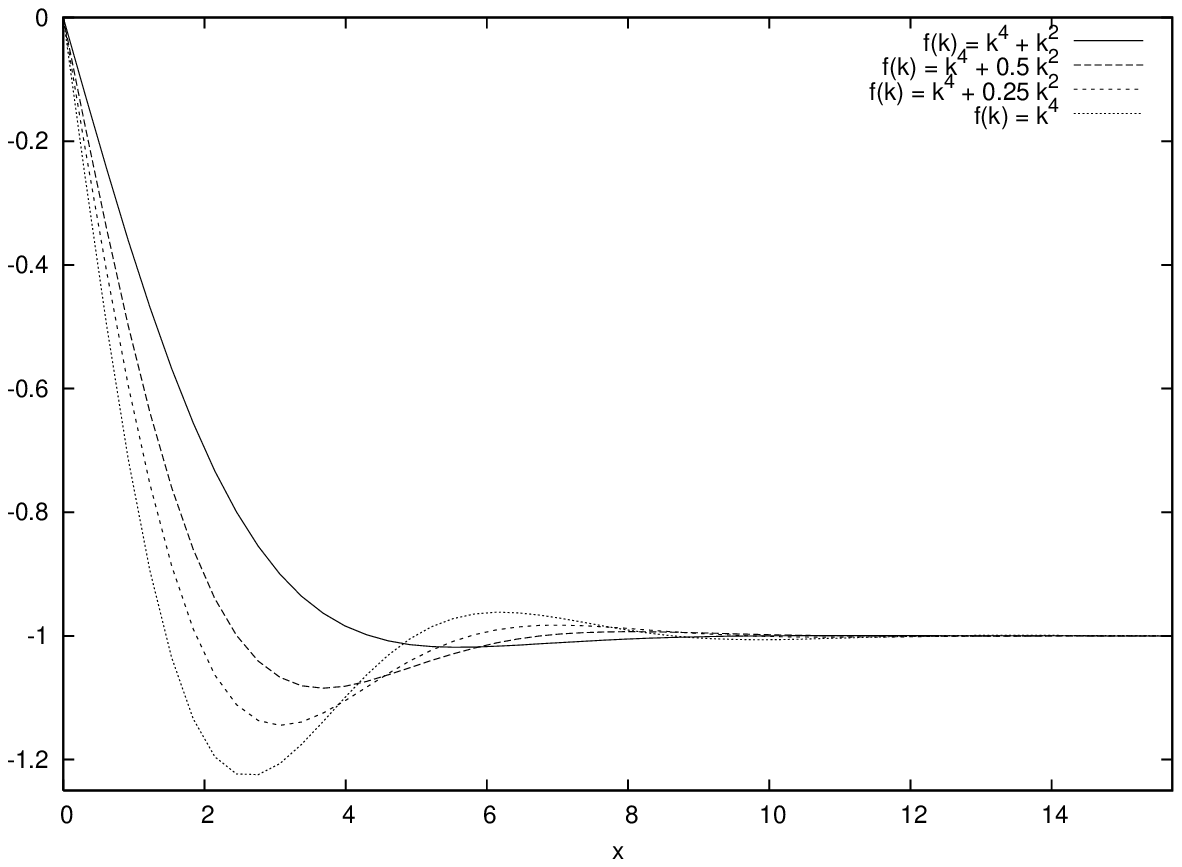} 
   \caption{
   		Numerical simulations of Eq.~(\ref{e:innerBurgers1})
		with the dissipation term $f(k) = k^4 + a k^2$, 
		with resolution $N = 1024$ and domain size $ L = 400 \pi $ for
		(a) and $L = 100 \pi $ for (b).  In (a) we represent
		the spectrum of the inner solution $ \vert \hat{u}^{(\mathrm{in})} \vert ^2 $ for $a = 1$ 
		(solid line), $ a = \textstyle\frac{1}{2}  $ (dashed line), 
		$ a = \textstyle\frac{1}{4}  $ (dash-dotted line)
		and $a = 0 $ (dotted line). In the inset we represent $ f(k) \vert \hat{u} ^{(\mathrm{in} )} \vert ^2 $
		for different values of $a$: $a = 1$  (solid line), $ a = \textstyle\frac{1}{2}  $ 
		(dashed line), $ a = \textstyle\frac{1}{4}  $ (dash-dotted line)
		and $a = 0 $ (dotted line). In (b) we represent the solutions $ u^{(\mathrm{in} )} $ in
		the physical space for $a = 1$  (solid line), $ a = \textstyle\frac{1}{2}  $ 
		(dashed line), $ a = \textstyle\frac{1}{4}  $ (dash-dotted line)
		and $a = 0 $ (dotted line). The exponentially decaying oscillations around $-1$ become 
		stronger for smaller $a$.
	   	}
   \label{fig:dissipation1}
\end{figure*}

Consider for example a dissipation term given by $f (k) = k^4 + a k^2 $. Then $ 
k_{\mathrm{i} } $ and $ k_{\mathrm{d} } $ can be calculated explicitly
\begin{eqnarray}
\label{e:examplebottleneck1}
k_{\mathrm{i} } (a) &=& \frac{1}{2} \,\sqrt [3]{2\,\pi +2\,\sqrt {16\,{a}^{3}+{\pi }^{2}}}-2\,{\frac {a
}{\sqrt [3]{2\,\pi +2\,\sqrt {16\,{a}^{3}+{\pi }^{2}}}}} \nonumber
\\
k_{\mathrm{d} } (a) &=& \frac{1}{2} \,\sqrt {-2\,a+2\,\sqrt {{a}^{2}+4}}  .
\end{eqnarray}
It follows that for $ a < a_{\star } \approx 0.2681736$,  where $ a_{\star } $ is the solution
of $ k_{\mathrm{i} } (a_{\star } ) = k_{\mathrm{d} } ( a_{\star } ) $, the consistency condition is 
violated and a significant contribution to the integral~(\ref{e:integral2}) has to come from
the intermediate (bottleneck) range.We perform detailed numerical simulations to confirm this result. 
In Fig.~(\ref{fig:dissipation1}) solutions of~(\ref{e:innerBurgers1}) are represented for $ a = 0, 1/4, 1/2 , 1 $ and a bottleneck is observed 
only for $ a = 1/4$ and $ a = 0$. For $ a = 1 $ there is clearly no bottleneck and there is 
practically no bottleneck in the case $ a = 1/2 $ either. 

We remark that  for $a \in ( 0 , 2^{ \frac{2}{3} }  3) $, that is for all values of $a$ that we analyzed above, 
Eq.~(\ref{e:zeta}) has complex solutions. Thus, the corresponding solutions of Eq.~(\ref{e:innerBurgers1}) 
oscillate around $\pm 1 $ for $  X \to \pm \infty $. But, as can be easily seen in
Fig.~(\ref{fig:dissipation1}) the amplitude of the oscillations decreases with increasing $a$. Thus, 
we view the oscillations appearing in the solutions when $f(k)$ falls off too fast with $ k \to 0 $
as another manifestation of the bottleneck phenomenon. The mere possibility of
oscillations in solutions of Eq.~(\ref{e:innerBurgers1}) does not necessarily lead to a bump in
the spectrum.

\subsection{Perturbative expansion for the hyperviscous boundary-layer Burgers equation}
\label{s:perturbative}

A special case in which the bottleneck effect can be analyzed analytically is Eq.~(\ref{e:innerBurgers1}) 
with dissipation given by the function $ f(k) = k^{2 \alpha } $, when
$ \alpha $ is close to unity. We write
$ \alpha = 1 + \varepsilon $ and use $\varepsilon $ as a small parameter.
Noting that 
\begin{equation}
k^{2 \alpha } = k^2 \, k^{2 \varepsilon }  = k^2 \, \sum_{n=0}^{\infty } 
\frac{1}{n!} \, ( 2 \ln k )^n \, \varepsilon ^n ,
\end{equation}
and assuming that $u^{\mathrm{in} } $ has an expansion in powers of $\varepsilon $
\begin{equation}
\label{e:leading-asymptoticsBLHBurgersExpansion}
u^{\mathrm{in} } = \sum_{n=0}^{\infty } u^{(n)}  \varepsilon ^n ,
\end{equation}
with $u^{(0)} = - \tanh \frac{x}{2} $ being the exact solution of Eq.~(\ref{e:innerBurgers1}) 
in the case $ f(k) = k^2 $, we obtain the following system of 
equations for the functions $u^{(n)} $, $n\geq 1$: at the leading order $n=1$
\begin{equation}
(2 \ln \sqrt{-\partial _x^2 } ) ( - \partial _x^2 ) u_0 +
( - \partial _x^2 ) u^{(1)}  + u_0 \partial _x u^{(1)} +
u^{(1)} \partial _x u_0  = 0 ,
\end{equation} 
and for $ n > 1 $
\begin{equation}
\hspace*{-2.6cm}
( - \partial _x^2 ) u^{(n)}  + u_0 \partial _x u^{(n)} +
u^{(n)} \partial _x u_0 + 
\sum_{m=1}^{n} 
\frac{ ( 2 \ln \sqrt{ - \partial _x^2 } )^m }{m!} ( - \partial _x^2 )
u^{(n-m)} + \sum_{m=1}^{n-1} u^{(m)} \partial _x u^{(n-m)} = 0.
\end{equation}
Now we show that at every fixed $n$, in particular at $n=1$, the function $u^{(n)}$
can be explicitly written in terms of $u^{(m)}$, with $ 0 \leq m < n$ and their 
Fourier transforms. Indeed, upon integrating the above equations we 
can rewrite them as
\begin{equation}
\label{e:perturbativeInhLinEqs}
\partial _x u^{(n)} = u_0 u^{(n)} - g^{(n)} ,
\end{equation}
where, for $n > 1$ 
\begin{equation}
\hspace*{-1.5cm}
g^{(n)} = \frac{2^n}{n!} \partial _x \, (\ln \sqrt{ - \partial _x^2 } )^n 
u_0 + \sum_{m=1}^{n-1} \frac{2^m}{m!} \partial _x \,  
(\ln \sqrt{ - \partial _x^2 } )^m u^{(n-m)} - \frac{1}{2} 
\sum_{m=1}^{n-1} u^{(m)} u^{(n-m)} 
\end{equation}
and 
\begin{equation}
g^{(1)} = 2 \partial _x \, \ln \sqrt{ - \partial _x^2 } u_0 ,
\end{equation}
for $n=1$.
Since for all $n \geq 1$ functions $u^{(n)} $ are odd,
we can write the solutions of the linear inhomogeneous equations 
(\ref{e:perturbativeInhLinEqs}) as
\begin{equation}
u^{(n)} (x) = - \frac{1}{\cosh ^2 \frac{x}{2} } \, 
\int_0^x g^{(n)} (x^{\prime } ) \cosh ^2 \frac{x^{\prime } }{2} \, 
dx^{\prime } ,
\end{equation}
or, inserting the expressions for $g^{(n)}$, as
\begin{eqnarray}
u^{(n)} = 
&-& \sum_{m=1}^{n} \frac{2^m }{m!} \, \frac{1}{\cosh ^2 \frac{x}{2} } \,
\int_0^x  \Bigl( \partial _x  (\ln \sqrt{ - \partial _x^2 } )^m 
u^{(n-m)} \Bigr) \cosh ^2 \frac{x^{\prime } }{2} \, 
dx^{\prime } \nonumber \\ &+&
\frac{1}{2} \sum_{m=1}^{n-1} \, \frac{1}{\cosh ^2 \frac{x}{2} } \,
\int_0^x  u^{(m)} u^{(n-m)}  \cosh ^2 \frac{x^{\prime } }{2} \, 
dx^{\prime }.
\end{eqnarray}
Finally, by using explicit representation for the action of the 
pseudo-differential operators 
$ (\ln \sqrt{ - \partial _x^2 } )^m $ we obtain 
the following expression for the function $u^{(n)}$
\begin{equation}
u^{(n)} = 
\sum_{m=1}^{n} \frac{2^m }{m!} \, a^{(m)}_n (x)
+
\frac{1}{2} \sum_{m=1}^{n-1} \, b^{(m)}_n (x) ,
\end{equation}
where
\begin{eqnarray}
a^{(m)}_n &=& \frac{1}{\sqrt{2\pi } i } \, 
\biggl[ \int_{\mathbb{R} } \frac{ k^2 (\ln \vert k \vert )^m }{1 + k^2 }
(\mathcal{F} u^{(n-m)} ) (k) \, \sin kx \, dk \nonumber \\ &+& 
\frac{d}{dx} \tanh \frac{x}{2}  \int_{\mathbb{R} } 
( \ln \vert k \vert )^m (\mathcal{F} u^{(n-m)} ) (k) 
\frac{ \sin k x }{1 + k^2 } \, dk \biggr]
\end{eqnarray}
and
\begin{equation}
b^{(m)}_n = \frac{1}{\cosh ^2 \frac{x}{2} } \,
\int_0^x u^{(m)} (x^{\prime } ) u^{(n-m)} (x^{\prime } )  
\, \cosh ^2 \frac{x^{\prime } }{2} \, dx^{\prime } .
\end{equation}
To study the bottleneck effect, it is enough to consider the first
order term in Eq.~(\ref{e:leading-asymptoticsBLHBurgersExpansion}),
which gives
\begin{equation}
\hspace*{-1.4cm}
u^{(1)} (x) =  2 \, \int_{\mathbb{R} } \, 
\frac{k^2 \ln \vert k \vert }{ \sinh (\pi k) } \, 
\frac{\sin k x }{1 + k^2 } \, dk + 
\frac{d}{dx} \, 
2 \, \tanh \frac{x}{2} \, 
\int_{\mathbb{R} } \, 
\frac{\ln \vert k \vert }{ \sinh (\pi k) } \, 
\frac{\sin k x }{1 + k^2 } \, dk 
\end{equation}
or, in the Fourier space,
$$
\left( \mathcal{F} u^{(1)} \right) (k) = 
\frac{2 \, \sqrt{2 \pi } }{i}  \frac{k^2 \, 
\ln \vert k \vert }{1 + k^2 } \, \frac{1}{ \sinh \pi k } + 
\frac{2 \, \sqrt{2 \pi } }{i} \, k \, 
\int_{\mathbb{R} } \frac{1}{(k^{\prime } )^2 + 1} \, 
\frac{\ln \vert k^{\prime } \vert }{ \sinh \pi k^{\prime } } \,
\frac{1}{\sinh \pi (k - k^{\prime } ) } \, dk^{\prime }, $$
where the integral has to be regularized in a suitable sense.

\subsection{Truncated solutions}
\label{s:truncated}

The arguments presented in the previous section imply that  
the best way to generate a bottleneck is to take for $f(k)$ a function which vanishes for $k$ smaller than a 
certain cut-off (which, without any loss of generality, we take to be $1$) and is infinite for $k$ above the cut-off
\begin{equation}
f_{\mathrm{tr} } (k) = \left\{ \begin{array}{lcc} 0           & \mathrm{for} & \vert k \vert < 1 ,\\
								    + \infty & \mathrm{for} & \vert k \vert > 1 .
								    \end{array} \right. 
\end{equation}
However, it is not clear how to implement such a dissipation term in Eq.~(\ref{e:innerBurgers1}).
We approximate such a cut-off dissipation term by considering a function $f(k)$ which depends on a certain
parameter in such a way that when the parameter tends to infinity, $f(k)$ tends to 
$f_{\mathrm{tr} } (k) $. 
Here we consider two examples of such functions: (i) hyperviscosity
$ f(k) = k^{2 \alpha } $ in the limit $\alpha \to \infty $  , a problem which 
has also been studied in \cite{Boyd94} and (ii)  a cosh-dissipation term exponentially growing for
$ \vert k \vert \to \infty $ 
\begin{equation}
\label{e:cosh1}
f(k) = e^{ - \mu }  ( \cosh \mu k - 1) ,
\end{equation}
introduced in \cite{Bardosetal} and studied further in  \cite{ZhT}. Both functions tend to 
$f_{\mathrm{tr} } (k) $ for $\alpha \to \infty $ and $\mu \to \infty $ but behave differently 
in the dissipation range, as we have seen in Section~\ref{s:difference-differential}.

\begin{figure*}[htbp]
   \centering
   \includegraphics[scale=0.6]{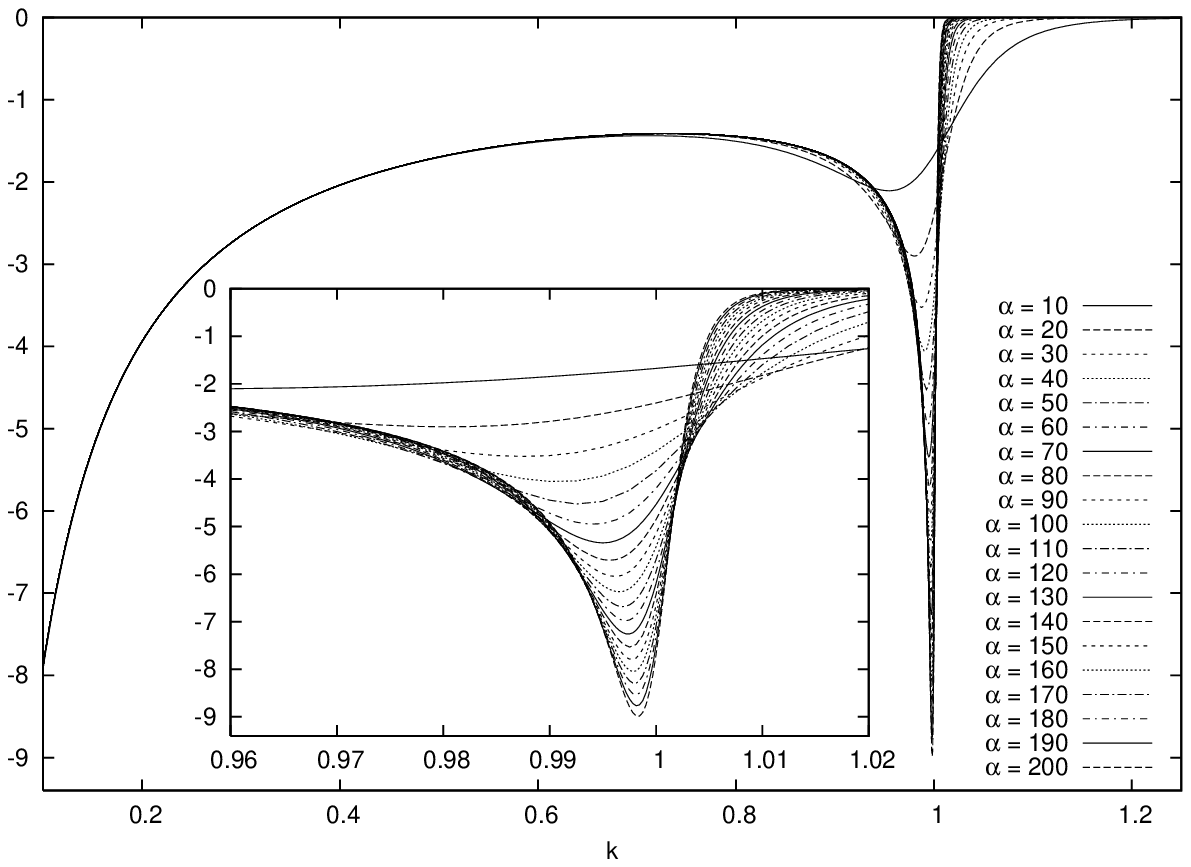} 
   \includegraphics[scale=0.6]{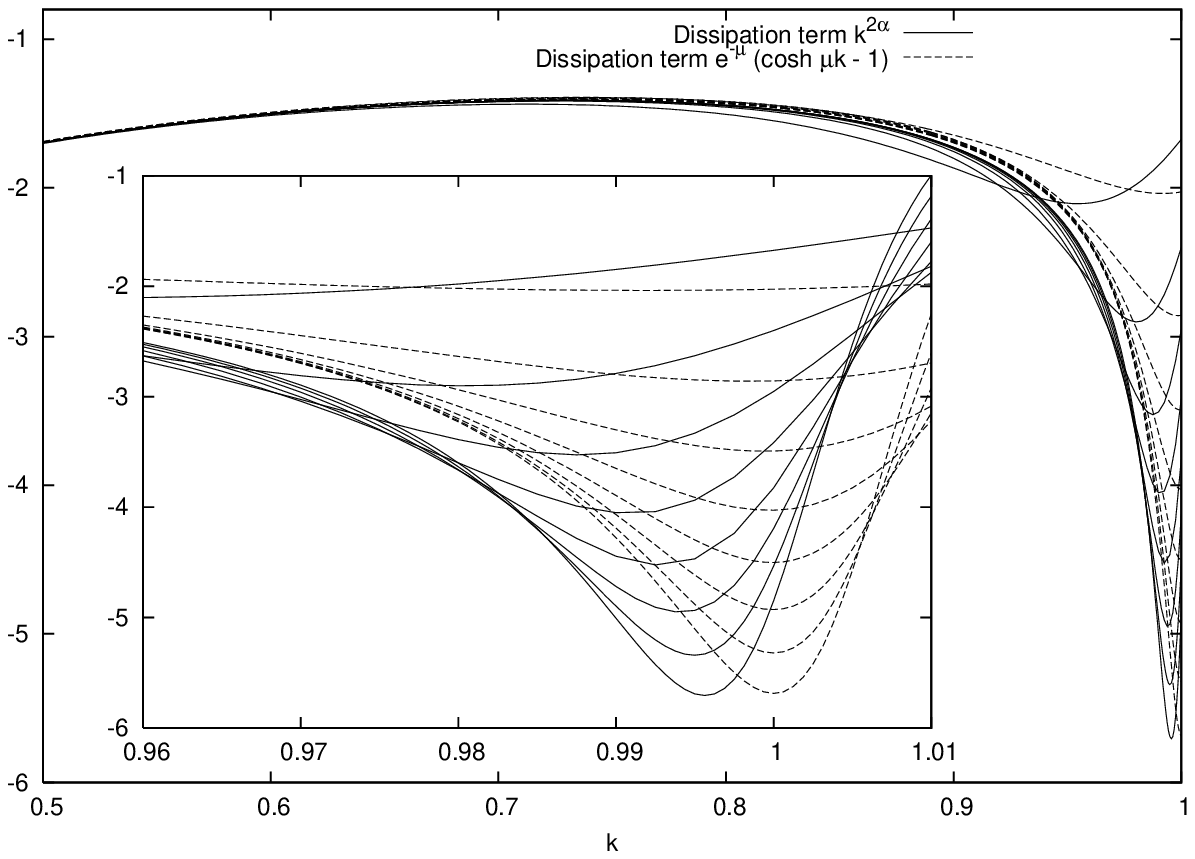} 
   \caption{
	   Numerical simulations of Eq.~(\ref{e:innerBurgers1})
		with hyperviscous dissipation terms  and cosh-dissipation
		terms.  In (a) we represent solutions for hyperviscous 
		dissipation terms with $ \alpha = 10, 20,...,200$. With increasing $\alpha $
		the solution tends to zero for $k>1$ and seems to acquire a well-defined 
		limit for $k<1$. In (b) we compare solutions for hyperviscous dissipation
		terms with $\alpha = 10, 20, ..., 80$ (solid lines) and cosh-dissipation terms with
		$\mu = 20, 40, ..., 160$ (dashed lines). 
	   	}
   \label{f:dissipation2}
\end{figure*}

For both types of dissipation we found that the solutions in the Fourier space seem to tend
to a well-defined limit for $ \vert k \vert < 1 $ and tend to zero for $ \vert k \vert > 1 $; this is illustrated 
in Fig.~(\ref{f:dissipation2}). The latter 
observation follows immediately from Eq.~(\ref{e:integral2}). 
The former follows from the numerical results for the hyperviscous and the cosh-dissipation terms, 
with representative plots shown in Fig.~(\ref{f:dissipation2}), which also suggest that the limiting function, which we denote
by $ u_{\infty } $, does not depend on the precise form of $f(k)$. Numerically, 
the high $\alpha $ and $\mu $ solutions in 
the neighborhood of $\pm 1$ are well described by the functional form
\begin{equation}
\label{e:infinity1}
\hat{u} _{\infty } (k) = \left\{ 
\begin{array}{lcc}
a (1 + k)^{- \Delta } + b , & - 1 < k  < 0, & k \sim -1 ,
\\
- a (1 - k  )^{ - \Delta } - b , & 0 < k  < 1 & k \sim 1 .
\end{array}
\right.
\end{equation}
A good agreement of the numerical data with the functional form~(\ref{e:infinity1}) is achieved for 
$\Delta \approx 2/3$ and $ a \approx 0.2 $, $b \approx 0.8 $.

Unfortunately, we did not manage 
to establish an equation for $ u_{\infty } $, and thus, we do not have a theory which would explain
the exponent $ \textstyle\frac{2}{3} $ in its Fourier space representation.  
The main difficulty in establishing such an equation consists in determining the $ \alpha \to \infty $
or $\mu \to \infty $ limit of the right hand side of Eq.~(\ref{e:innerBurgers1}) which we 
denote by $ R(u_{\infty } ) $. 
Clearly, whereas the support of the limiting function itself is $ \mathrm{supp} \, u_{\infty }  = [-1,1] $, 
the support of $R(u_{\infty } )$ is contained 
in  $ ( - \infty , - 1] \cup [ 1 , + \infty )$. More precisely, since on the left hand side of Eq.~(\ref{e:innerBurgers1}) we have a quadratic term, the support is equal to 
$ [ - 2, - 1] \cup [ 1 , 2]$. Some information about $R(u_{\infty } )$ can be obtained by using 
Eq.~(\ref{e:integral1})  for $ g(u) = \textstyle\frac{1}{2} u \vert u \vert $ and 
$G(u) = \mathrm{sign} (u) $, so that
\begin{equation}
\label{e:truncestimate1}
\int_{\mathbb{R} }  \mathrm{sign} ( u^{( \mathrm{in} ) } ) f  \left( \sqrt{ - \frac{d^2}{dX^2} }  \right) 
 u^{( \mathrm{in} ) }  \, dX = 1 .
\end{equation}
Based on numerical results we assume that $  \mathrm{sign} ( u^{( \mathrm{in} ) } ) = 
- \mathrm{sign} ( X ) $, which gives us the following relation for the term on the
right hand side of Eq.~(\ref{e:innerBurgers1})
\begin{equation}
\label{truncestimate2}
\int_{\mathbb{R} }  \mathrm{sign} ( X ) f  \left( \sqrt{ - \frac{d^2}{dX^2} }  \right) 
 u^{( \mathrm{in} ) }  (X) \, dX = - 1 ,
\end{equation}
from which follows, via Parseval's theorem,
\begin{equation}
\label{truncestimate3}
\int_{\mathbb{R} } \frac{f(k)}{k} \, \hat{u} ^{(\mathrm{in} )} (k) \, dk = - i \sqrt{\frac{\pi }{2} } .
\end{equation}
Relations~(\ref{e:integral2}) and~(\ref{truncestimate3}), combined with numerical results,  
suggest that $R(u_{\infty } )$ is a function and not a distribution. However, 
we did not succeed in determining the functional form of this function.

From the numerically obtained functional form of $\hat{u} _{\infty } $ we deduce 
the asymptotic form of $u_{\infty } $ in the physical space for $X \to \infty $
\begin{equation}
\label{e:truncatedBLHBurgers1}
u_{\infty } (X) \simeq \left\{ 
\begin{array}{lcc}
1 + a \sqrt{ \frac{2}{\pi } }  \, \Gamma \Bigl( \frac{1}{3} \Bigr) \, 
(-X)^{ - \frac{1}{3} }  \sin \Bigl( X + \frac{\pi }{6} \Bigr) 
& \mathrm{for}   & X \to - \infty ,
\\
- 1  + a \sqrt{ \frac{2}{\pi } }  \, \Gamma \Bigl( \frac{1}{3} \Bigr) \, 
X^{ - \frac{1}{3} }  \sin \Bigl( X - \frac{\pi }{6} \Bigr)  
& \mathrm{for}   & X \to + \infty .
\end{array}
\right.
\end{equation}

\section{Spectrum of the one-dimensional Burgers equation with modified dissipation}
\label{s:fullsolutions}

In Sections~\ref{s:smallReynolds} and \ref{s:bottlesection} we have studied simplified 
models derived for solutions of the Burgers equation. Whereas the results of Section~\ref{s:smallReynolds}
concern dissipation scales only,  Section~\ref{s:bottlesection} deals with the intermediate 
range between the inertial range and the dissipation range. 
In this section we shall see how far the results of the previous two sections can be used to analyze 
numerical solutions of the Burgers equation with modified dissipation in the 
Fourier space. We employ the following strategy: we solve the Burgers equation by 
using  high-precision pseudo-spectral simulations with the {\sl mpfun}-package \cite{Bailey}. 
This  approach allows us to analyze the solutions deep in the
dissipation range which becomes the more important the faster the dissipation term grows with $k$.
We employ the Exponential Time Differencing Runge-Kutta scheme 
\cite{ETDRK,Matthews}.

We concentrate essentially on the behavior
of solutions in two ranges: the dissipation range (or the high wavenumber range) and 
the bottleneck range (or the transition range from the inertial to the dissipation range). 

The functional form of solutions in the Fourier space in the
dissipation range is studied by using the example of  two different kinds of dissipation:
the hyperviscous dissipation and the cosh-dissipation. Here we have the
advantage that our numerical investigations of the $\vert k \vert \to \infty $
asymptotics  can be checked against the theoretical predictions of
Eqns.~(\ref{e:hyperviscous1}) and (\ref{e:cosh-dissipation1}).  This is
important in particular with regard to numerical studies of more general
equations, such as the Navier--Stokes equations, for which analytical results
concerning the form of the dissipation range are few. 

To analyze the asymptotics in the dissipation range numerically, we apply the 
asymptotic extrapolation procedure of van der Hoeven \cite{Hoeven09}. This procedure can be viewed 
as a sequence of transformation techniques in which the main idea is to remove the higher leading-order 
terms by applying a suitable sequence of transformation and then, knowing the sub-leading
order terms, to obtain the leading-order terms. The choice of the order and the type of transformations 
depends on the functional form of the analyzed sequence. In our case we essentially
take the sequence used in \cite{PF} and \cite{Bardosetal}.

For the hyperviscous dissipation term $ \nu ^{2 \alpha - 1} k^{2 \alpha } $ the 
dissipation range begins roughly at $ 1 / \nu $. Taking $ \nu $ to be of order one gives us a 
solution which lies entirely in the dissipation range. For such a solution the small Reynolds 
number results of Section~\ref{s:smallReynolds} apply in the first place and thus give us a 
means to check the validity of the small-Reynolds number expansion of Section \ref{s:smallReynolds}.
Unfortunately,  numerical analysis in the case of $\nu \sim 1 $ turns
out to be difficult, because for such values of $\nu $ the solution in the Fourier space
falls off very quickly, so that very high precision and extremely small time steps 
are required: the higher are the modes whose Fourier coefficients we calculate, 
the higher the precision and computational accuracy is needed.

\begin{figure*}[htbp]
   \centering
      \includegraphics[scale=0.6]{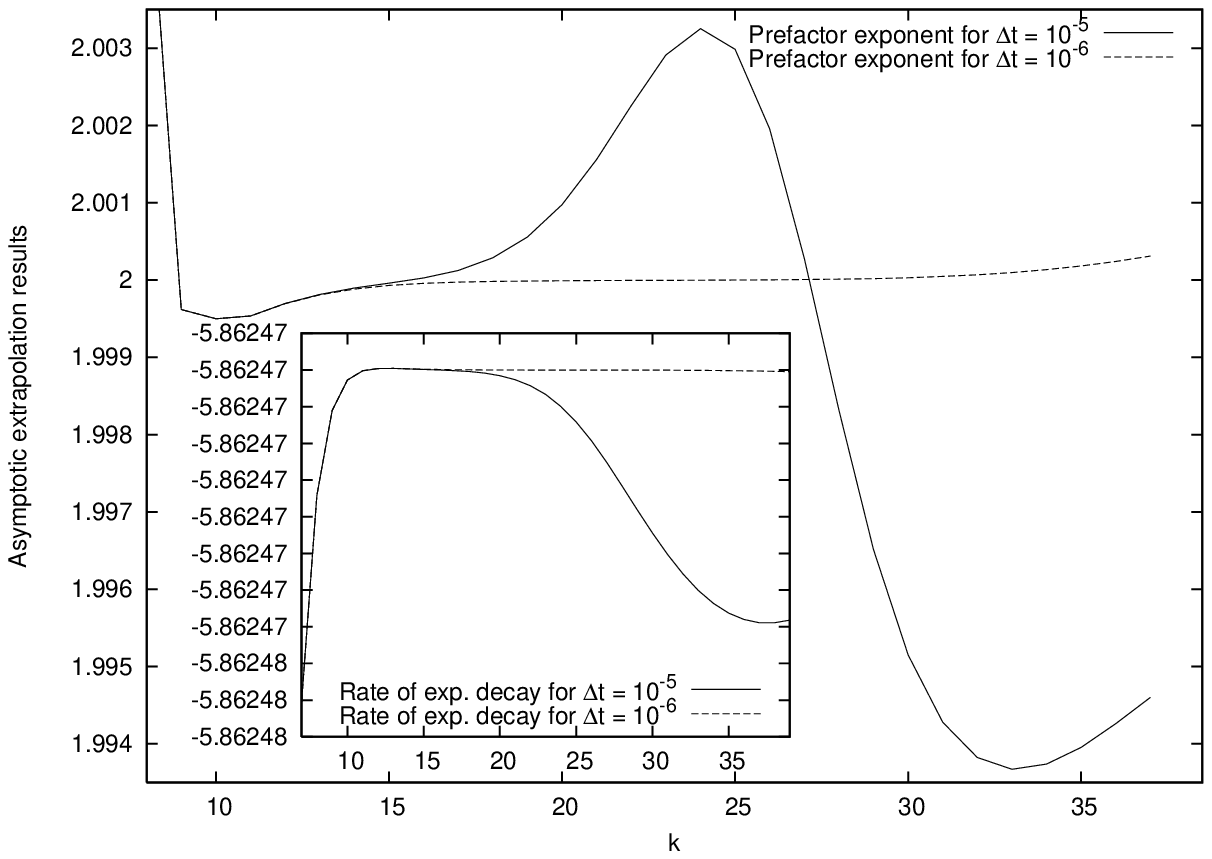} 
      \includegraphics[scale=0.6]{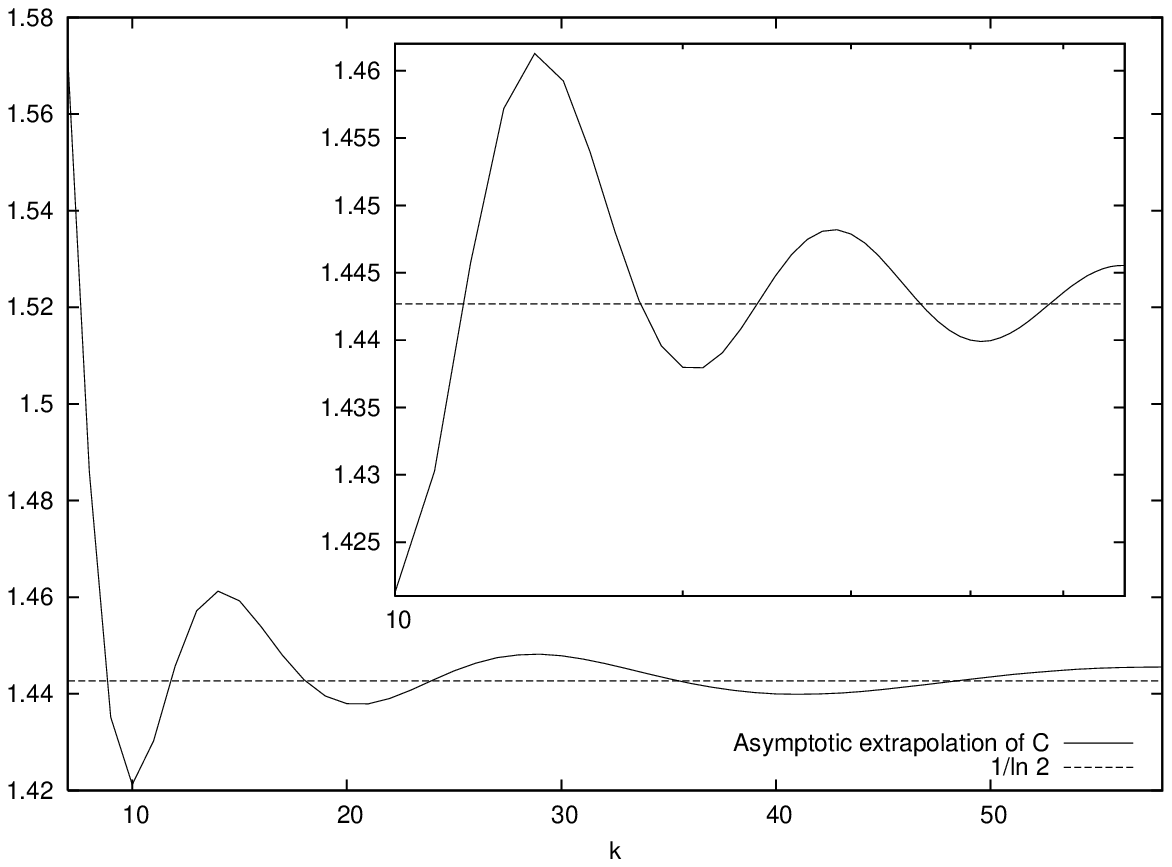}
   \caption{  
   		(a) Results of asymptotic extrapolation procedure applied to the Fourier coefficients of the solution
		of Eq.~(\ref{e:Burgers2}) with $\nu = 1$ and $\alpha = 2$, initial condition $\sin x $, calculated with 
		200 digits and time step $10^{-6}$. At the fourth stage of asymptotic extrapolation the sequence tends
		to the constant value $ - 2/\alpha $. The deviations from this value is of the order $10^{-4} $.
	   	}
   \label{f:high-precision1}
\end{figure*}

As a consequence, in the expansion~(\ref{e:hyperviscous1}) only the
leading and the two sub-leading terms can be reliably determined. For 
example, the exponent of the algebraic prefactor is obtained 
with a relative precision of order $10^{-4} $ whereas for the rate of exponential decay we get a precision of 
$10^{-7} $ as shown in Fig.~(\ref{f:high-precision1}).
We did not succeed in determining any further sub-leading terms, such as complex powers
of $k$, because of several reasons related to insufficiently small time steps and a lack of sufficient 
number of modes for extrapolation.

Simulations employing cosh-dissipation give results similar to the hyperviscous case. Hence for  the 
dissipation term $f(k) = ( \cosh k - 1 ) $ the leading-order term $ \exp ( - C k \ln k ) $
can be clearly identified. In particular, the numerical value of the constant $C =  - 1 / \ln 2 $ conjectured
in \cite{Bardosetal} and predicted by Eq.~(\ref{e:cosh-dissipation1}) can be confirmed with certainty
as shown in Fig.~(\ref{f:high-precision1}b).
Unfortunately, the determination of higher order term in the asymptotic expansion is 
hampered by the logarithmic-scale oscillations present in the next-order correction 
$ \exp (  k g ( \ln k ) ) $ (function $g(\cdot ) $ is periodic with period $\ln 2 $) giving rise to the
logarithmic scale oscillations in Fig.~(\ref{f:high-precision1}b).

What happens when the dissipation term starts acting at wavenumbers much higher than one, so that
a substantial inertial range can be developed? As we shall see now,  although the functional form 
predicted by the small Reynolds number expansion can be identified in the dissipation range, 
for dissipation terms producing large bottlenecks this becomes increasingly difficult, since one
has to go to higher and higher wavenumbers to recover the asymptotic behavior of the Fourier coefficients
of solutions. At the same time the parts of the bottleneck region adjacent to the inertial range 
are satisfactorily described by the linear asymptotic approximation based on Eq.~(\ref{e:leading-asymptoticsBLHBurgers1}).

\begin{figure*}[htbp]
   \centering
      \includegraphics[scale=0.6]{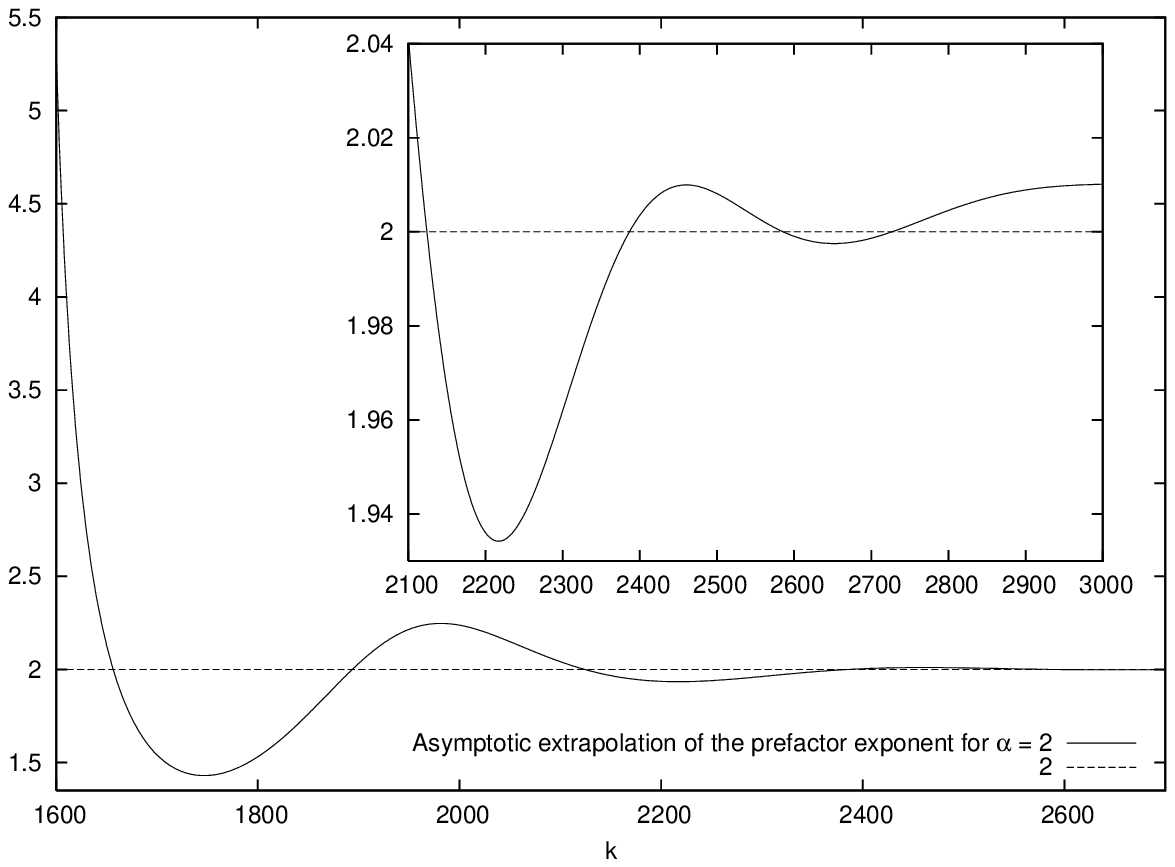} 
      \includegraphics[scale=0.6]{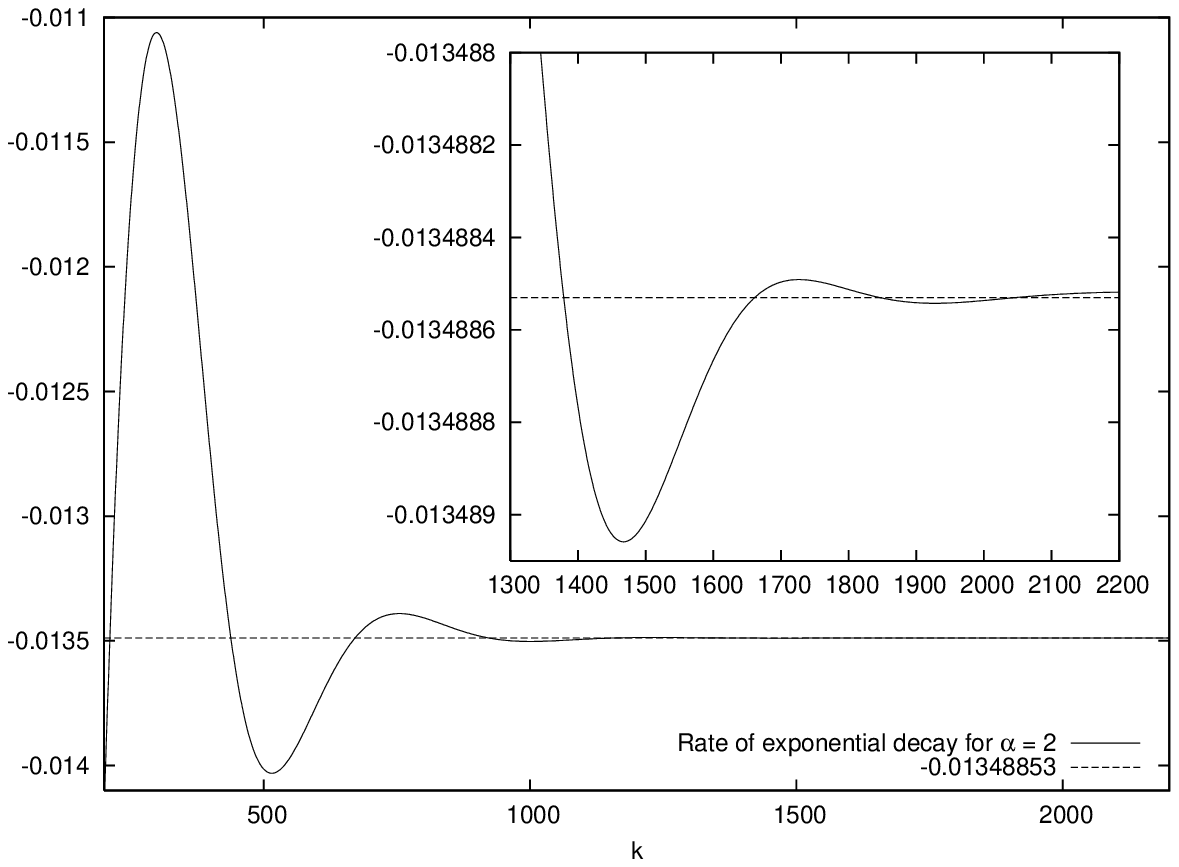}
   \caption{  
   		Results of asymptotic extrapolation procedure applied to Fourier coefficients of the solution
		of Eq.~(\ref{e:Burgers2})with $\nu = 10^{-8} $ and $\alpha = 2$, initial condition $\sin x $, calculated with 
		54 digits, time step $\Delta t = 10^{-5}$ and resolution $N = 2^{13} $ at time $t = 1.1$. 
		Panel (a) are the results for the algebraic prefactor exponent for which the analytical 
		value is $2 \alpha - 2 = 2$. The deviations from the theoretical value are of order $10^{-2} $ for 
		wavenumbers  between $2400$ and $2800$.
		Panel (b) shows the results of asymptotic extrapolation for the rate of exponential decay
		in the leading-order term.
	   	}
   \label{f:high-precision2}
\end{figure*}

We have calculated solutions of the Burgers equation with hyperviscous dissipation terms
of the type  $\nu ^{2 \alpha - 1} k^{2\alpha } $ for small $\nu $ numerically by using high-precision for several values of
$\alpha $. For $ \alpha = 2$ the functional form of the dissipation range is identified quite accurately: For 
example, for the numerical solution using $\nu ^3 = 10^{-8} $, the 
exponent of the algebraic prefactor is determined with the relative precision of the order $10^{-2} $, for the rate
of exponential decay the relative precision is of the order $10^{-5} $ as shown in Fig.~(\ref{f:high-precision2}).
Remarkably, to obtain the functional form of the solution in the dissipation 
range accurately even for $\alpha = 2$ we have to go quite far beyond the wavenumber $ 1 / \nu   
\approx 464 $, i.e.  the wavenumber at which dissipation sets in.  For example, as can be seen in 
Figure~\ref{f:high-precision2}, the relative error in the
determination of the prepfactor exponent drops below $10^{-2} $ only for $ k > 5/ \nu  \approx 2321 $. 
For $ \alpha = 3$ and $ \nu ^5 = 10^{-14} $ we would have to go even farther beyond the wavenumber 
$ 1/\nu ^{ \frac{1}{5} } \approx 631 $: for $ k > 5/\nu ^{ \frac{1}{5} } \approx 3155 $ and up to $N/2 = 10^{12} $ the 
asymptotic extrapolation procedure for the algebraic prefactor exponent does not converge to any value,
displaying oscillations similar to those in Fig.~(\ref{f:high-precision2}a), but much stronger.
For the rate of exponential decay the error is of the order $10^{-4} $ if we assume that the algebraic prefactor is
$k^4$. 
Even worse convergence to asymptotic behavior is observed for the exponentially growing 
dissipation terms for which even the identification of the leading order term requires resolutions much higher
than $ 1/ \nu $.

\begin{figure}[htbp]
   \centering
      \includegraphics[scale=0.8]{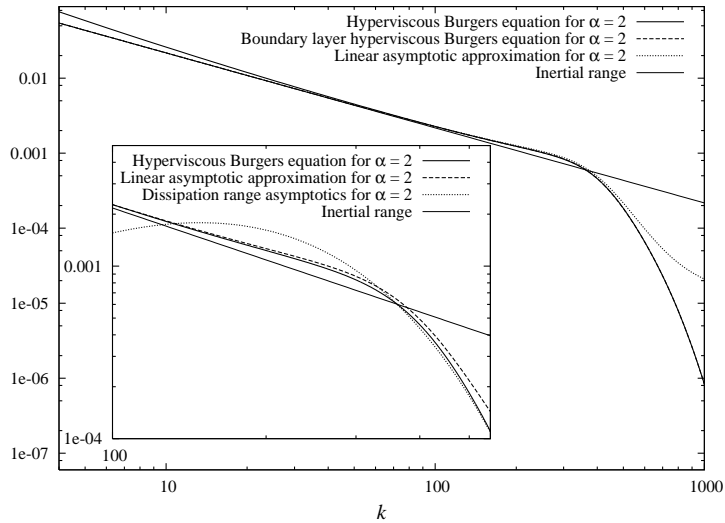} 
   \caption{  
   		Log-log scale representation of the Fourier coefficients of the	solution
		of Eq.~(\ref{e:Burgers2}) with $\nu = 10^{-8} $ and $\alpha = 2$, initial condition $\sin x $, 
		calculated with 54 digits, time step $\Delta t = 10^{-5}$ and resolution $N = 2^{13} $ at time $t = 1.1$. 
		We compare this solution with the solution of the boundary layer Burgers equation, 
		the rescaled linearized asymptotic solution
		and the inertial range scaling $ \sim k^{-1} $. In the inset the numerical solution 
		of Eq.~(\ref{e:Burgers2}) is compared with the functional form in the dissipation range, the rescaled linearized 
		asymptotic solution and the inertial range scaling.
	   	}
   \label{f:bottleneckok}
\end{figure}

For the bottleneck region we use the results of Section~\ref{s:bottleoscillations}, in particular the numerical
values of the amplitude $ A $ and of the phase shift $\phi $. The functional form of solutions in the bottleneck 
region (Eq.~(\ref{e:leading-asymptoticsBLHBurgers1})) is approximated by the linearized asymptotic solution 
$ \hat{u} ^{(\mathrm{in} )}_{\mathrm{as} } (k) $ of the boundary layer Burgers equation~(\ref{e:innerBurgers1}) as 
\begin{equation}
\label{e:rescaleBLHB}
\hat{u} (k) \simeq \frac{1}{\sqrt{2 \pi } }  \, \frac{J}{k_{\mathrm{e} } }  \,  
\hat{u} ^{(\mathrm{in} )}_{\mathrm{as} } (k/k_{\mathrm{e} } ) , 
\end{equation}
where $J$ is the jump in the entropic solution at the shock, by Fast-Legendre transforms, and 
$k_{\mathrm{e} } = \sqrt[ 2 \alpha - 1]{J}  /\nu $ is the effective dissipation wavenumber. As can be seen
in Fig.~(\ref{f:bottleneckok}) on the example of the hyperviscous Burgers equation with $\alpha = 2 $
and $\nu ^3 = 10^{-8} $, the agreement of the approximative solution with the actual solution is extremely good.

\section{Conclusions}

In this article we have seen by using the example of the
one-dimensional Burgers equation with a modified 
dissipation term how the structure 
of solutions of a hydrodynamical equation can be 
described by simplified models which 
can be obtained from the original equation
by systematic reduction.  We have concentrated on 
the far dissipation region and the transition region
from the inertial range to the dissipation range.

To study the far dissipation region we have 
presented  a method which allows us to 
study solutions of hydrodynamical equations at 
small Reynolds numbers in domains with 
periodic initial conditions. This method takes
advantage of the fact that for initial conditions 
with suitably restricted modes the interaction between
modes is restricted  and solutions
can be obtained recursively without any errors 
due to truncation or time-stepping. It is applicable 
to more general hydrodynamical equations such as the 
Navier--Stokes equations which was one of the reasons 
to present it here.  For the one-dimensional 
Burgers equation in the limit of long
times the problem can be simplified even further, so that the 
problem reduces to a non-linear difference-differential equation. 
By using this equation we have studied the 
high wavenumber asymptotics in detail and verified the results by 
using high-precision pseudo-spectral numerical simulations. 

We have seen that the transition range from the inertial range to the dissipation 
range in the case of the Burgers equation 
can be described quite well by a linearized solution of the boundary
layer problem in the neighborhood of shocks. However, 
in contrast to the study of the far dissipation range where the analysis 
has been done by a method which a priori does not use any special 
properties of the one-dimensional Burgers equation, in the study of the 
intermediate range we had to rely on a very special property of the 
Burgers equation.
A further 
drawback is that we were not able to determine analytically 
the amplitude and the phase of oscillations near the shocks and had to use
numerics to determine them. 

How far the analysis presented in this article applicable to 
the Navier--Stokes equations? As we stated above, the method 
for the analysis of the far dissipation range presented here can be extended to 
the Navier--Stokes equations in arbitrary dimensions. The main difference to the 
Burgers case is that the corresponding recursion relations are hard to deal with
analytically and have to be studied numerically using high-precision arithmetics,
analogously to singularities of the Euler equation \cite{MBF05,PMFB06,P10}.
The results of this ongoing work will be published elsewhere. 

The treatment of the bottleneck problem seems to be more difficult 
because the Burgers type analysis does not apply to incompressible flows. 
It is known that the bump in the energy spectrum appears together with oscillations 
in the physical space \cite{banerjee14,SPF}, but we are still far from understanding this 
phenomenon analytically. 

\bigskip

We thank D. Banerjee for several useful discussions and collaborations; we also 
thank D. Vincenzi for encouragement.  SSR
acknowledges DST (India) project ECR/2015/000361 for financial support.

\bigskip

\end{document}